\documentclass[twocolumn]{aastex7}
\usepackage{siunitx}
\usepackage{amsmath,amssymb}
\usepackage{verbatim}
\usepackage{color}
\usepackage{hyperref}
\usepackage{booktabs}
\usepackage{float}
\usepackage{needspace}
\usepackage{graphicx}
\usepackage{enumitem}
\usepackage{halloweenmath}
\usepackage{natbib}
\usepackage{mathrsfs}
\usepackage{enumitem}
\usepackage{graphicx}  
\usepackage{adjustbox} 
\usepackage{rotating}
\usepackage[english]{babel} % To obtain English text with the blindtext package
\usepackage{blindtext}
\definecolor{linkcolor}{rgb}{0,0,0.25}
\hypersetup{
  colorlinks=true,        % false: boxed links; true: colored links
  linkcolor=linkcolor,    % color of internal links
  citecolor=linkcolor,    % color of links to bibliography
  filecolor=linkcolor,    % color of file links
  urlcolor=linkcolor      % color of external links
}

\newcounter{address}
\defcitealias{Heiles_2003}{HT03}
\defcitealias{Wilcots}{WM}

\newcommand{\NRAO}{\affiliation{National Radio Astronomy Observatory, 520 Edgemont Road, Charlottesville, VA 22903, USA}}

\DeclareMathAlphabet{\mathsc}{OT1}{cmr}{m}{sc}
\def\testbx{bx}%
\DeclareRobustCommand{\donen}[2]{%
  \relax\ifmmode
  \ifx\testbx\f@series
    {\mathbf{#1\,\mathsc{#2}}}\else
    {\mathrm{#1\,\mathsc{#2}}}\fi
  \else\textup{#1\,{\mdseries\textsc{#2}}}%
  \fi
}
\usepackage[table]{xcolor}
\usepackage[dvipsnames]{xcolor}

\begin{document}

\title{Probing Cold Gas in IC~10 with Neutral Hydrogen Absorption}

\author[0000-0002-4110-8769]{Ioana~A.~Stelea}
\email{stelea@wisc.edu}
\affiliation{University of Wisconsin–Madison, Department of Astronomy, 475 N Charter St, Madison, WI 53703, USA}

\author[0000-0002-3418-7817]{Sne{\v z}ana Stanimirovi{\'c}}
\email{sstanimi@wisc.edu}
\affiliation{University of Wisconsin–Madison, Department of Astronomy, 475 N Charter St, Madison, WI 53703, USA}

\author[0000-0001-9504-7386]{Nickolas~M.~Pingel}
\email{nmpingel@iu.edu}
\affiliation{University of Wisconsin–Madison, Department of Astronomy, 475 N Charter St, Madison, WI 53703, USA}
\affiliation{Department of Astronomy, Indiana University, 727 East Third Street, Bloomington, IN 47405, USA}

\author[0009-0005-1781-5665]{Hongxing Chen}
\email{hchen792@wisc.edu}
\affiliation{University of Wisconsin–Madison, Department of Astronomy, 475 N Charter St, Madison, WI 53703, USA}

\author[0000-0001-9605-780X]{Eric W. Koch}
\email{koch.eric@gmail.com}
\affiliation{National Radio Astronomy Observatory, 800 Bradbury SE, Suite 235, Albuquerque, NM 87106 USA}

\author[0000-0002-2545-1700]{Adam~K.~Leroy}
\email{leroy.42@osu.edu}
\affiliation{Department of Physics, The Ohio State University, 191 W. Woodruff Avenue, Columbus, OH 43210, USA}

\author[0000-0002-5204-2259]{Erik Rosolowsky}
\email{rosolowsky@ualberta.ca}
\affiliation{Department of Physics, 4-183 CCIS, University of Alberta, Edmonton, AB T6G 2E1, Canada}

\author[0000-0003-2896-3725]{Chang-Goo Kim}
\email{changgookim@gmail.com}
\affiliation{Department of Astrophysical Sciences, Princeton University, Princeton, NJ 08544, USA}

\author[0000-0002-5480-5686]{Alberto~D.~Bolatto}
\email{bolatto@umd.edu}
\affiliation{Department of Astronomy, University of Maryland, College Park, MD 20742, USA}
\affiliation{Space Telescope Science Institute, 3700 San Martin Drive, Baltimore, MD 21218, USA}

\author[0000-0002-1264-2006]{Julianne J.~Dalcanton}
\email{jdalcanton@flatironinstitute.org}
\affiliation{Astronomy Department, University of Washington, Box 351580, Seattle, WA 98195-1580, USA}
\affiliation{Center for Computational Astrophysics, Flatiron Institute, 162 Fifth Avenue, New York, NY 10010, USA}

\author[0000-0003-4961-6511]{Michael P. Busch}
\email{mpbusch@nrao.edu}
\altaffiliation{Jansky Fellow of the National Radio Astronomy Observatory}
\NRAO

\author[0009-0007-7949-6633]{Harrisen Corbould}
\email{harrisen@ualberta.ca}
\affiliation{Department of Physics, 4-183 CCIS, University of Alberta, Edmonton, AB T6G 2E1, Canada}

\author[0000-0003-0235-3347]{J. R. Dawson}
\email{joanne.dawson@mq.edu.au}
\affiliation{School of Mathematical \& Physical Sciences and Astrophysics \& Space Technologies Research Centre, Macquarie University, NSW 2109, Australia}
\affiliation{Australia Telescope National Facility, CSIRO Space \& Astronomy, PO Box 76, Epping, NSW 1710, Australia}

\author[0000-0002-1185-2810]{Cosima Eibensteiner}
\email{ceibenst@nrao.edu}
\altaffiliation{Jansky Fellow of the National Radio Astronomy Observatory}
\NRAO

\author[0000-0002-3227-4917]{Amanda Kepley}
\email{akepley@nrao.edu}
\affiliation{National Radio Astronomy Observatory, 520 Edgemont Road, Charlottesville, VA 22903, USA}

\author[]{Melanie Krips}
\email{krips@iram.fr}
\affiliation{IRAM, Domaine Universitaire, 300 rue de la Piscine, 38406 Saint-Martin-d’Hères, France}

\author[0000-0002-7743-8129]{Claire~E.~Murray}
\email{clairemurray56@gmail.com}
\affiliation{Space Telescope Science Institute, 3700 San Martin Drive, Baltimore, MD 21218, USA}

\author[0000-0001-6326-7069]{Julia Roman-Duval}
\email{duval@stsci.edu}
\affiliation{Space Telescope Science Institute, 3700 San Martin Drive, Baltimore, MD 21218, USA}

\author[0000-0003-3351-6831]{Daniel R. Rybarczyk}
\email{rybarczyk@astro.wisc.edu}
\affiliation{University of Wisconsin–Madison, Department of Astronomy, 475 N Charter St, Madison, WI 53703, USA}

\author[0000-0003-0605-8732]{Evan~D.~Skillman}
\email{skill001@umn.edu}
\affiliation{Minnesota Institute for Astrophysics, School of Physics and Astronomy, University of Minnesota, 116 Church Street SE, Minneapolis, MN 55455, USA}

\author[0000-0003-1356-1096]{Elizabeth Tarantino}
\email{etarantino@stsci.edu}
\affiliation{Space Telescope Science Institute, 3700 San Martin Drive, Baltimore, MD 21218, USA}

\author[0000-0002-5877-379X]{Vicente Villanueva}
\email{vicente.avl365@gmail.com}
\affiliation{Instituto de Estudios Astrof{\'i}sicos, Facultad de Ingenier{\'i}a y Ciencias, Universidad Diego Portales, Av.\ Ej{\'e}rcito Libertador 441, 8370191 Santiago, Chile}
\affiliation{Millennium Nucleus for Galaxies, MINGAL}

% \affiliation{Departamento de Astronom{\'i}a, Universidad de Concepci{\'o}n, Barrio Universitario, Concepci{\'o}n, Chile}

% Instituto de Estudios Astrofísicos, Facultad de Ingeniería y Ciencias, Universidad Diego Portales, Av. Ejército Libertador 441, 8370191 Santiago, Chile

\author[0000-0002-0012-2142]{Thomas~G.~Williams}
\email{thomas.williams@physics.ox.ac.uk}
\affiliation{Sub-department of Astrophysics, Department of Physics, University of Oxford, Keble Road, Oxford OX1 3RH, UK}
\affiliation{UK ALMA Regional Centre Node, Jodrell Bank Centre for Astrophysics, Department of Physics and Astronomy, The University of Manchester, Oxford Road, Manchester M13 9PL, UK}

\title[]{The Local Group L-band Survey: Probing Cold Atomic Gas in IC~10 with Neutral Hydrogen Absorption}

\begin{abstract}
We present the first localized detections of the cold neutral medium (CNM) in IC~10, offering a rare view of dense atomic gas in a low-metallicity ($Z/Z_{\odot} \sim 0.27$) dwarf galaxy. As a low-metallicity starburst, IC~10's interstellar medium conditions could reflect small scale physics conditions that mirror those of early galaxies, providing a unique window into the heating and cooling processes that shaped the interstellar medium in early-Universe environments. Leveraging the high angular ($<5'' \sim 15$ pc) and spectral ($0.4$ km s$^{-1}$) resolution of the Local Group L-band Survey, we searched for HI absorption against nine continuum radio sources and detected absorption along three sightlines corresponding to internal radio emission sources within IC 10. Using Gaussian decomposition and radiative transfer, we characterize the CNM, deriving spin temperatures of $\sim 30$–$55$ K, column densities of $(0.6-3.0) \times10^{21}$ cm$^{-2}$, cold HI fractions of $\sim 21$–$37\%$, and line widths of $~\sim5.6-13.6$ km s$^{-1}$.
For each individual detection of HI absorption, we find corresponding molecular emission from $^{12}$CO ($J$ = 1–0), HCO$^+$ ($J$ = 1–0), and HCN ($J$ = 1–0) at similar velocities and with comparable linewidths, indicating a well-mixed cold atomic and molecular medium. In IC~10, the CNM shows a clear kinematic connection to the high-density ISM, implying a stronger dynamical coupling with molecular gas than in the Milky Way, in line with expectations for low-metallicity environments. At the $\sim$15 pc scales probed by slightly extended HII regions in IC10, unresolved CNM clouds likely contribute to line blending, so the observed broad HI linewidths may partly reflect spatial and kinematic averaging.
\end{abstract}      


\section{Introduction}\label{intro}
The interstellar medium (ISM) is a dynamic, multi-phase environment that plays a crucial role in the lifecycle of galaxies. Comprised of ionized, molecular, and neutral components, the gaseous phase of the ISM serves as a reservoir of fuel for star formation. Among its components, the atomic neutral hydrogen (HI) and its abundance is particularly important, as the transition from HI to molecular hydrogen (H$_2$) acts as a throttle in the interstellar gas cycle \citep{naomi}.

HI can exist in multiple phases \citep{Field1969, wolfire95, wolfire2003}, with the long-lived thermally stable phases being the warm neutral medium (WNM) and the cold neutral medium (CNM). Theoretical models for Solar neighborhood conditions predict kinetic temperatures ($T_k$) ranging from $25-250$ K for the CNM and $5000-8000$ K for the WNM \citep{wolfire2003, Bialy2019}. 
The CNM represents the primary reservoir of gas that transitions into molecular hydrogen \citep{Spitzer1975}.
Being more susceptible to gravitational collapse, the CNM also serves as a precursor to giant molecular clouds (GMCs) and stars \citep{heitsch2005, kimostriker2006, Dobbs_2014}. The fraction of HI in the CNM thus provides vital constraints on the formation of molecular clouds \citep{Spitzer1975, Elmegreen1993, Krumholz2009, Bialy2016}.
The local ISM conditions, such as metallicity, directly affect the abundance of the CNM and the physical conditions within it \citep{wolfire95, wolfire2003,Bialy2019, kim2024ApJ...972...67K}. As metallicity serves as a proxy for a galaxy's evolutionary stage, studying the CNM across a broad metallicity range is key to understanding its role in galactic evolution. The nearby low-metallicity galaxies are especially useful as they offer insights into CNM's behavior in high-redshift galactic environments.

The key CNM properties, e.g. excitation temperature of HI (spin temperature\footnote{At high densities the excitation temperature closely traces the kinetic temperature \citep{field58, liszt2001}.}, $T_s$) and the CNM fraction ($f_{\rm{CNM}}$), are best constrained observationally by observing HI absorption in the direction of background radio continuum sources. The HI absorption directly measures the HI optical depth ($\tau$) along the line of sight; however, an estimate of a nearby HI emission is also needed to constrain excitation temperature via radiative transfer equations \citep[e.g.,][]{Dickey2003, Heiles_2003}. 
To solve radiative transfer equations it is essential that both HI absorption and emission sample similar solid angles, which results in the need for high-resolution HI emission observations \citep[for discussion of how a miss-match between angular resolution of HI emission and absorption affects $T_s$ calculations see Section 4.2 in][]{21spongea}.

As HI absorption detections require bright background sources and sensitive observations,  there have only been a handful of HI absorption detections in external galaxies outside the Magellanic System, specifically M31, M33 \citep{dickeym31,dickeyboth, otjerm31} and M82 \citep{Wills1998}. 
In addition, matching of pencil-beam sharp optical depth spectra with high-resolution HI emission needed to solve radiative transfer equations can be problematic even at the distance of the Magellanic Clouds \citep[e.g.][]{Chen2025a}. As a result, characterizing CNM properties in external galaxies to investigate the dependence of the CNM fraction on metallicity has largely been limited to theoretical predictions \citep{wolfire95,kim2024ApJ...972...67K,Smith2023} and indirect observational constraints \citep[e.g. the use of $\text{[CII]}$ emission;][]{Herrera-Camus2017}.
The width of HI emission-line profiles has also been commonly used in external galaxies to separate cold from warm HI and study their spatial distribution \citep[e.g.][]{Braun1997, deBlok2006,  Tamburro2009,Mogotsi2016}.
However, the interpretation of HI emission profiles is usually complicated by limitations in the spatial/spectral
resolution  and overall degeneracy in the line broadening caused by turbulence and the blending of emission components arising from velocity crowding and a mix of thermal phases \citep{koch2021}

\textbf{The Local Group L-band Survey (LGLBS)} \citep{Koch2025} provides unprecedented spectral and angular resolution 21-cm and 1–2 GHz radio continuum data of six Local Group galaxies (M31, M33 and four star-forming dwarfs - IC~10, NGC 6822, IC 1613, and WLM). With angular resolutions of $<5 \arcsec$ for the 21-cm line (10–20 pc) and $<2 \arcsec $ for continuum imaging (5–10 pc), along with a velocity resolution of 0.4 km s$^{-1}$, LGLBS offers an excellent opportunity to study cold neutral gas structure at individual cloud scales. Using LGLBS data, \cite{nickngc} detected for the first time
five CNM components in NGC~6822 \citep{nickngc}. In this study, we search for cold HI in absorption in another LGLBS low-metallicity dwarf galaxy -- IC~10.

The paper is structured as follows. First we provide background information for IC~10 in Section \ref{sec:IC~10}. Then, in Section \ref{sec:obsv} we summarize key survey specifications and describe our analysis pipeline. In Section~\ref{sec:absp}, we present our sources and considerations for non-detections, while Section~\ref{sec:rt} details our radiative transfer methods. Section~\ref{sec:results} presents the derived physical properties of the CNM and compares them with observations of similar dwarf galaxies in the Local Group. Finally, in Section~\ref{sec:discussion}, we place our results in the broader context of IC~10’s complex, turbulent, star-forming ISM and discuss potential biases in our approach.

\section{Previous Work on IC~10}\label{sec:IC~10}
IC~10 is a well-studied irregular, low-metallicity dwarf galaxy in the Local Group, with a metallicity of $12 + \log{(\text{O/H})} = 8.37$ \citep{Cosens2024} \citep[$Z/Z_{\odot} \sim 0.27$;][]{garnett1990, margini2009, Karachentsev2004, Karachentsev2013AJ....145..101K}. Located at 770 $\pm$ 100 kpc \citep{Sakai1999, DellAgli2018}\footnote{The distance to IC~10 is uncertain due to differential extinction at low Galactic latitudes. The reported value is from \cite{DellAgli2018} and is an intermediate value between distances used by LITTLE THINGS, \cite{Sanna2008}, \cite{Gerbrandt2015}, and \cite{Putman2021}. See \cite{Koch2025} for more information.}, IC~10 is classified as a starburst galaxy based on its rich population of Wolf-Rayet (WR) stars \citep{Massey1995, stellarpop2002} and elevated values of star formation rate (SFR) for its mass and metallicity. SFR estimates range from 0.2 \citep{Kennicutt1994} to 0.6 M$_{\odot}\, \text{yr}^{-1}$ \citep{Yang1993, Borissova2000, GilDePaz2003} depending on the assumed extinction toward the galaxy. These exceed the SFR of the more massive Small Magellanic Cloud (SMC) (0.05 M$_{\odot} \text{yr}^{-1}$ \citealt{Wilke2004}) and approach that of M33 ($0.45 \pm 0.10$ M$_{\odot} \text{yr}^{-1}$ over the past 100 Myr \citealt{Verley2009}). With its proximity, elevated SFR, and central starburst, IC~10 presents a unique opportunity to study the properties of the CNM in a low-metallicity, high-feedback environment.

The HI distribution in IC~10 is markedly irregular, with disrupted kinematics that point to a turbulent evolutionary history \citep[e.g.,][]{Shostak1989, Wilcots, littlethings}. 
Two dominant scenarios have been proposed to explain the disturbed HI kinematics. First, IC~10 may be interacting or merging with an undetected companion \citep{nidever, ashley}, or it could be affected by a close passage to M~31 based on proper motion measurements \citep{Brunthaler2007,Bennet2024}. Second, stellar winds and supernovae resulting from the intense recent star formation are shaping the HI morphology \citep{Wilcots}. 

A defining feature of IC~10’s disrupted morphology is the presence of several large holes in its HI disk \citep{Shostak1989, Wilcots, holes2020}. These cavities, requiring on the order of $10^{50}$–$10^{51}$ erg to form \citep{Wilcots}, highlight the impact of intense stellar feedback in the galaxy’s ongoing starburst. Both stellar winds from young massive stars and supernovae likely contribute to their formation and expansion. The star-formation induced feedback is reflected in IC~10’s rich population of HII regions \citep{hodgelee} and young stellar clusters ($\leq10$ Myr) concentrated in H$\alpha$-emitting regions \citep{Vacca2007, Sanna2009, Yin2010}, whose energetic outputs drive the disruption and shaping of the surrounding atomic gas. 
\section{Observations and Data Processing}\label{sec:obsv}
HI observations of IC~10 used in this study were obtained with the VLA as part of the LGLBS collaboration. Detailed information on the observing strategy, data calibration, quality assurance, and correction for short spacings using observations with the Green Bank Telescope are provided in \cite{Koch2025}. The final imaging was performed on the University of Wisconsin–Madison Center for High Throughput Computing using the method developed by \cite{Pingel2024}.

\begin{figure*}
    \centering
    \includegraphics[width=0.8\linewidth]{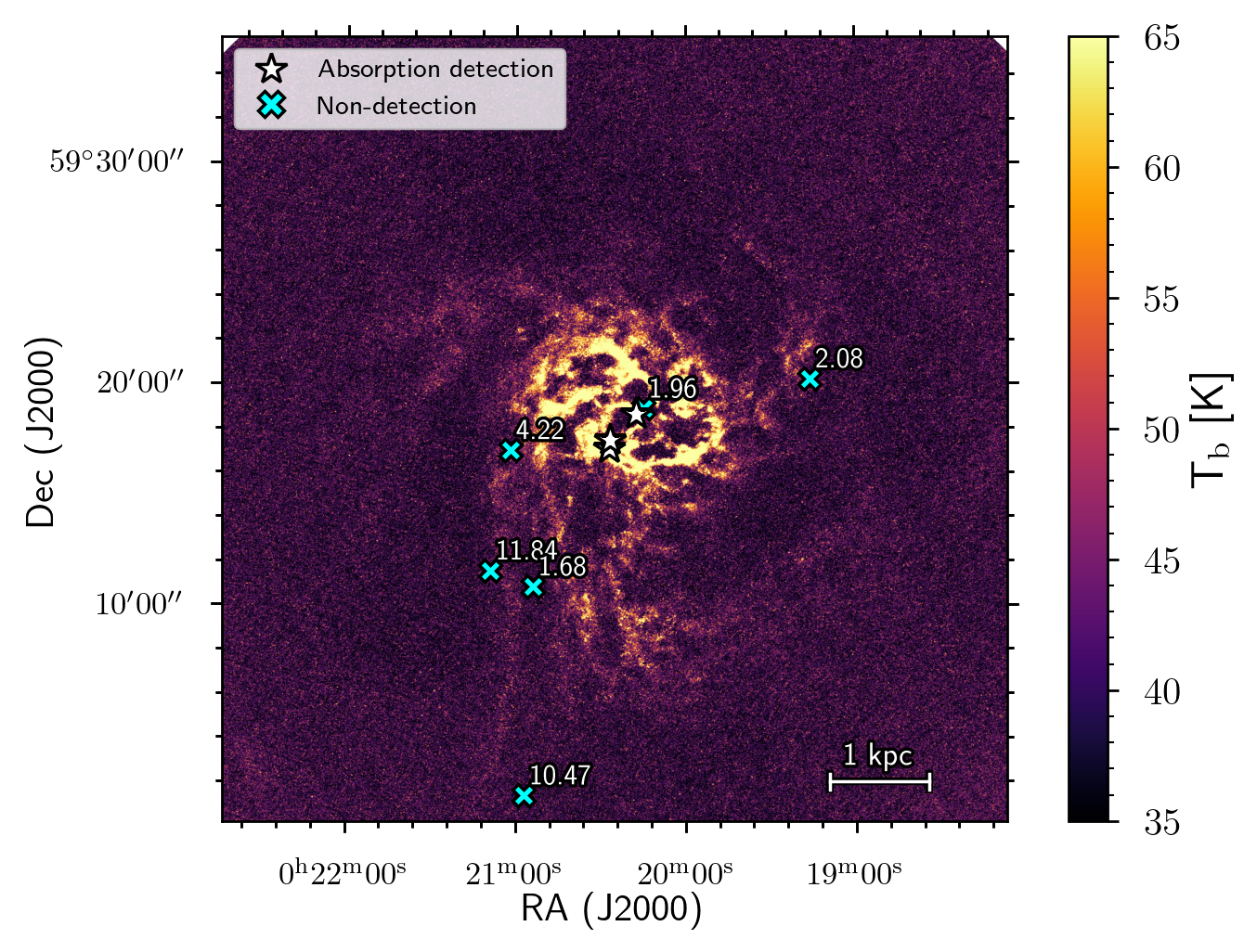}
    \caption{Peak HI brightness temperature map of IC~10. The 9 radio continuum sources analyzed for HI absorption, a mix of internal and background sources, are marked: 3 internal sources with detected absorption are shown as stars, and 6 non-detections (internal or background) as crosses. For each non-detection, the integrated flux density (in mJy) is labeled. Integrated flux densities for all sources, including the detections, are listed in Table~\ref{tab:observations}.}
    \label{fig:peak}
\end{figure*}
The resulting ABCD+GBT HI emission cube has a synthesized beam size of 4.22$''$ × 3.98$''$, corresponding to $\approx$14.5–15.4 pc at the distance of $770\pm100$ kpc to IC~10. This is comparable to the spatial resolution achieved by observations of the nearer Magellanic Clouds \citep[part of GASKAP-HI,][]{DickeyGASKAP}.

The root-mean square (RMS) noise of the ABCD+GBT HI emission cube is 23 K (corresponding to 0.6 mJy beam$^{-1}$)  per 0.4 km s$^{-1}$ velocity channel. Assuming optically thin emission, we reach 5$\sigma$ column density sensitivity of $4.2 \times 10^{20}~\mathrm{cm^{-2}}$, considering a velocity linewidth of 10 km s$^{-1}$. Our sensitivity is three times better than the sensitivity achieved in previous VLA D-configuration observations by \citet{Wilcots}, and similar to the sensitivity reached by the LITTLE THINGS survey which utilized VLA BCD configurations \citet{littlethings}.\footnote{After smoothing our data cube to the \citet{Wilcots} resolution ($11''$, 12.2 km s$^{-1}$), our 5$\sigma$ sensitivity is $5.6 \times 10^{18}\ \mathrm{cm^{-2}}$, nearly three times better than their $1.6 \times 10^{19}\ \mathrm{cm^{-2}}$. When matched to \citet{littlethings} ($8.4''$, 2.5 km s$^{-1}$), we reach $2.1 \times 10^{19}\ \mathrm{cm^{-2}}$, consistent with their $2.35 \times 10^{19}\ \mathrm{cm^{-2}}$.} 
The key advantage of our LGLBS observations is higher angular resolution, achieved with the addition of VLA A-configuration observations, and high velocity resolution (0.4 km s$^{-1}$). 

\subsection{Absorption and Emission Extraction in the Direction of Radio Continuum Sources}\label{sec:absemsextract}

We follow, with minor modifications, the HI absorption spectrum extraction procedure from \citet{nickngc}, which builds on the method of \citet{Dempsey}. 

\underline{\textit{Narrowband-continuum imaging:}} We first produce a narrow-band continuum image by resampling all continuum visibilities to a common spectral axis and reference frame, and combine data into a single measurement set. The visibilities are reweighted based on noise measurements in channels free of emission. 
We then create a narrow-band continuum image using spectral line data from the A and B configurations by performing a multi-frequency synthesis imaging over a 4 MHz bandwidth centered on 1420.6946 MHz (corresponding to the 21-cm line), while excluding foreground Milky Way emission. This bandwidth is chosen to ensure the synthesized continuum closely matches the line data, while remaining sensitive enough for strong continuum detections in the 0.4 km s$^{-1}$ channels of the spectral cube.
We apply 20,000 Hogbom clean iterations to produce the narrow-band continuum image with a $3.26'' \times 2.36''$ beam size and $0.75'' \times 0.75''$ pixel size. This image is used to search for discrete radio continuum sources. 

\underline{\textit{Selection of target continuum sources:}} We identified 49 candidate continuum sources for the HI absorption searching the narrowband continuum image, using the automated source-finding package \texttt{aegean} \citep{aegean1, aegean2} with default parameters. The resulting catalogue contains source positions, peak flux densities ($S_{\text{peak}}$), integrated flux densities ($S_{\rm{int}}$) and morphological information (major/minor axes and position angles). These sources are either compact objects within IC~10 (e.g., HII regions or supernova remnants) or background radio sources (AGN, radio galaxies) dominated by synchrotron emission. The latter appear as unresolved point sources, while the former present modest spatial extent.

For all 49 candidates, we generated spectral-line sub-cubes ("cubelets" hereinafter) to search for HI absorption. One such cubelet has a synthesised beam of $4.92'' \,\, \times4.58''$. To further characterize the candidates, we ran \texttt{aegean} on these cubelets and identified sources with flux densities ranging from 0.2 to 10.67 mJy beam$^{-1}$.
Based on the measured RMS noise in the cubelets \footnote{$\delta S = 0.41$ mJy beam$^{-1}$ per 0.4 km s$^{-1}$ channel}, the effective sensitivity requires continuum flux densities of at least $\sim 3$ mJy beam$^{-1}$ to detect absorption with optical depth $\tau \sim 0.5$, comparable to values reported in the SMC \citep{Dempsey}. Accordingly, in Table~\ref{tab:observations} we list the nine brightest sources (those that provide the best sensitivity for CNM detection) together with their spatial extents as measured from the spectral-line cubelets. These sources are a mix of external background radio sources and internal compact sources of radio emission within IC10.

Furthermore, to ensure completeness, all 49 sightlines were visually inspected in \texttt{CARTA}. This inspection revealed that absorption features could still be identified toward fainter sources, leading us to adopt a practical flux threshold of $S \geq 1.67$ mJy beam$^{-1}$ (optical depth RMS $\sim 0.33$), sufficient for detecting strong absorption features (e.g., $\tau > 1$).

\underline{\textit{Absorption line extraction:}} We extract HI absorption spectra from cubelets centered on the positions of selected radio continuum sources. Each cubelet is a spectral-line subcube with a $50'' \times 50''$ field of view. Like in the case of imaging NGC6822  \citep{nickngc} we exclude baselines shorter than 1km during the spectral-line imaging for IC10.
The narrow-band continuum image retains all baselines to ensure accurate detection of compact sources. From each cubelet, we extract the HI absorption spectrum in the direction of the continuum source. To isolate the relevant pixels, we use the elliptical fits from \texttt{aegean} to define a mask that selects pixels spatially associated with the continuum emission. In line-free velocity channels ($-475$ km s$^{-1} \leq v_{\text{LSRK}} \leq -230$ km s$^{-1}$), we compute the mean brightness temperature ($T_B$) along the line of sight for each pixel associated with the continuum source. The final absorption spectrum is obtained by summing over these pixels in each spectral channel, weighting by the square of the corresponding line-of-sight mean $T_B$.

\underline{\textit{Emission line extraction:}} For the purpose of radiative transfer calculations, we extract the HI emission spectrum along each absorption sightline from the continuum-subtracted A+B+C+D emission cube. For each source, we construct the final HI emission spectrum by measuring the mean $T_B$ within an annulus centered on the source, spanning from $4''$ to $8''$ in radius (approximately 1–2 synthesized beam widths). This approach minimizes contamination from strong absorption at the source position, ensuring that the resulting emission profile reflects the unabsorbed HI along the same line of sight. However, we note that a subset of the background sources exhibit modest spatial extension. For these cases, we conducted manual checks by redefining the radius of the inner annulus based on the size of the source diameter rather than the synthesized beam. These tests showed no significant differences in the shape or brightness temperature of the resulting spectra.

\subsection{Noise Estimates for HI absorption and emission spectra}
We calculate the noise in the emission spectrum by spatially rebinning the cube, such that each pixel approximately corresponds to one synthesized beam in area. This way, we account for pixel correlation and obtain an independent measurement of the uncertainty. The uncertainty in the emission spectrum, $\sigma_T^{\text{em}}$, is then calculated as the channel-by-channel standard deviation of $T_B$ within the annular region in the rebinned cube.

In the case of absorption we calculate the $1\sigma$ optical depth noise spectrum in units of $e^{-\tau}$ as outlined by \citet{roy2013}, with further details provided in \citet{nickngc}. Namely, we first measure the standard deviation in signal-free channels of the $e^{-\tau}$ spectrum, $\sigma_{\mathrm{cont}}$. To account for an increase in system temperature due to extended emission, we compute the average brightness temperature spectrum, $T_{\mathrm{em}}(v)$, from the ABCD emission cube. The resulting $1\sigma$ optical depth noise profile is given by: 
\begin{equation}\label{eq:1}
    \sigma_\tau(v)=\sigma_{\mathrm{cont}} \frac{T_{\mathrm{sys}}+\eta_{\mathrm{ant}} T_{\mathrm{em}}(v)}{T_{\mathrm{sys}}}
\end{equation}
where the system temperature $T_{\mathrm{sys}} = 25\,\mathrm{K}$ and the antenna efficiency $\eta_{\mathrm{ant}} = 0.35$, which accounts for the fraction of large-scale emission coupled to the antenna and contributing to the measured system noise. 

\begin{table*}[htb]
\centering
\setlength{\tabcolsep}{10pt}
\caption{Observed properties of candidate radio continuum sources, as measured from the spectral-line cubelets (with synthesised beam of $4.92'' \times 4.58''$). Columns include: (1) Right ascension (RA) in hms (J2000); (2) Declination in dms (J2000); (3) integrated continuum flux density ($S_{\text{int}}$) in mJy, with uncertainties; (4) source angular extent (minor and major axis, in arcsec) measured from the cubelets; (5) sensitivity to absorption ($\sigma_{e^{-\tau}}$) expressed in units of $e^{-\tau}$; (6) uncorrected optically thin HI column density along the line of sight (Eqn. \ref{eq:nhi,thin}); (7) Name of the source across different catalogues.}  
\label{tab:observations}
\resizebox{\textwidth}{!}{
\begin{tabular}{lcccccc}
\hline
\hline
Source  RA (hms) & Source DEC (dms)  & $S_{\text{int}}$ (mJy) & Source Size ($'' \, \times \, ''$)& $\sigma_{e^{-\tau}}$ & $N_{\text{HI, uncorr}}$ (cm$^{-2}$) & Notes \\
$0^{h}20^{m}27^{s}$ & $59^{\circ}17'06''$ & $3.99 \pm 0.45 $ &  $6.49'' \times 5.68''$ & 0.36 & 5.64 $\times10^{21}$ & \textsc{WBBH J002027+591706.1} \citep{Westcott2017} \\
$0^{h}20^{m}27^{s}$ & $59^{\circ}17'28''$ & $2.13 \pm 0.28$ & $4.94'' \times 4.45''$  & 0.26  & 3.02$\times10^{21}$ & \textsc{WBBH J002027+591728.7} \citep{Westcott2017} \\
$0^{h}20^{m}17^{s}$ & $59^{\circ}18'39''$ & $9.79 \pm 0.21$ & $5.72'' \times 5.04$ & 0.41 & 1.57$\times10^{21}$  & \textsc{WBBH J002017+591839.7}  \citep{Westcott2017} \\
\hline
$0^{h}20^{m}57^{s}$ & $59^{\circ}01'24''$  & $10.47 \pm 0.27$ & $4.91'' \times 4.52''$ &  0.41 & 5.47$\times10^{19}$  & \textsc{NVSS J002057+590122} \citep{nvss} \\
$0^{h}21^{m}02^{s}$  & $59^{\circ}17'01''$  & $4.22 \pm  0.24$  & $6.45'' \times 5.66''$
& 0.60 & 1.13$\times10^{21}$ & \textsc{NVSS J002100+591659} \citep{nvss}  \\
$0^{h}21^{m}09^{s}$  & $59^{\circ}11'34''$ & $11.84 \pm 0.21$ & $5.27'' \times 4.82''$ & 
0.20 & 2.78$\times10^{20}$ & \textsc{NVSS J002108+591132} \citep{nvss} \\
$0^{h}19^{m}16^{s}$ & $59^{\circ}20'15''$ & $2.08 \pm 0.17$ &  $4.91'' \times 4.48''$ & 
1.16 & 6.11$\times10^{20}$ & \textsc{NVSS J001914+592008} \citep{nvss} \\
$0^{h}20^{m}15^{s}$ & $59^{\circ}18'54''$ & $1.96 \pm 0.15$ & $4.79'' \times 4.62''$ & 0.70 & 5.63$\times10^{20}$ &  \textsc{WBBH J002015+591853.9} \citep{Westcott2017} \\
$0^{h}20^{m}54^{s}$ & $59^{\circ}10'51''$ & $1.68 \pm 0.15$ & $4.75''\times 4.44''$ & 0.80 &  4.43$\times10^{20}$ & \textsc{NVSS J002054+591101} \citep{nvss} \\
\hline
\end{tabular}
}
\end{table*}
\section{HI Absorption in IC~10}\label{sec:absp}
\begin{figure*}
    \centering
\includegraphics[width=\linewidth]{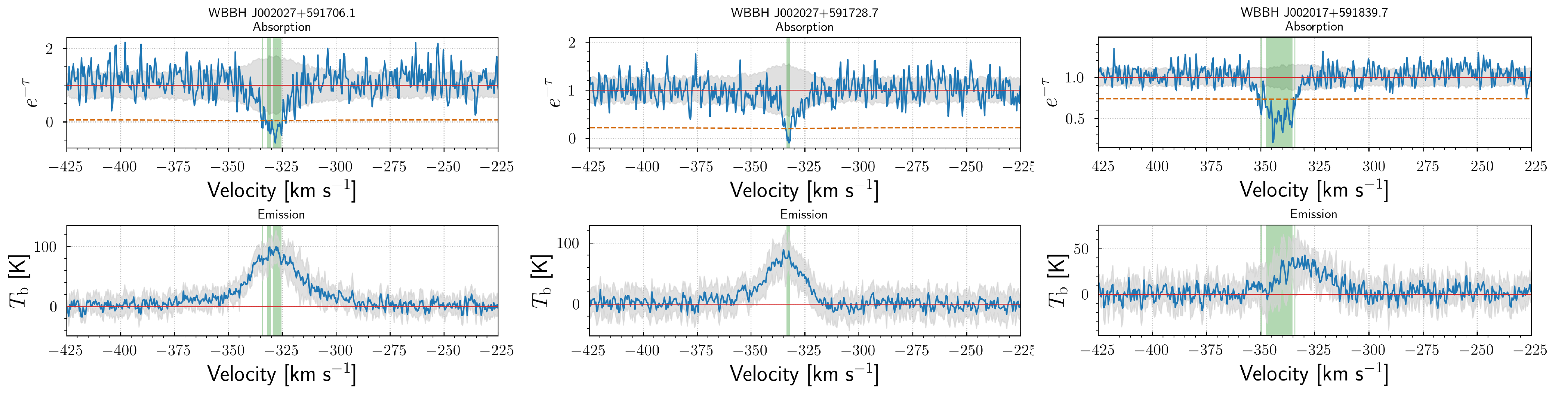}
\caption{Absorption (top) and emission (bottom) spectra for the three sightlines with detected HI absorption, constructed as described in Section~\ref{sec:absemsextract}. In both panels, the gray shaded region shows the $1\sigma$ noise level. In the absorption spectra, the red horizontal line indicates the $e^{-\tau}=1.0$ continuum level, and the dashed orange line marks the $3\sigma$ noise threshold. Green shaded regions highlight channels where HI absorption exceeds $3\sigma$.}
    \label{fig:speconly}
\end{figure*}

Figure~\ref{fig:peak} presents the peak HI brightness temperature distribution of IC~10 from LGLBS, highlighting detailed small-scale structure, including several prominent HI shells. Overlaid are nine radio continuum sources with flux density $> 1.67$ mJy beam$^{-1}$ in the direction in which we searched for HI absorption. These sources are a mix of both external background sources (5) and internal (4) radio emission sources. For each continuum source, we extracted absorption and emission spectra as outlined in Section \ref{sec:absemsextract}. The three sources with detected HI absorption are marked with stars, while directions with non-detections are shown with crosses. 

Table~\ref{tab:observations} gives the key properties of the continuum sources and the directions they probe. We include source names (for detections, JRA–DEC), coordinates (RA/DEC in hms/dms), integrated flux ($S_{\text{int}}$), 
and the estimated size (in arcseconds).
The noise level of the absorption spectra is 
($\sigma_{e^{-\tau}}$ is provided in units of $e^{-\tau}$). The HI column density ($N_{\text{HI, uncorr}}$) is calculated under the optically thin assumption as:
     \begin{equation}\label{eq:nhi,thin}
    N_{\text{HI, uncorr}} = 1.823 \times 10^{18} \int T_B (v) \,\,dv \text{ cm}^{-2}
\end{equation} 
In the last column of the table, we list the names of each source as they appear in different catalogs. Sources catalogued by \cite{Westcott2017} are considered internal, while those identified in the NVSS catalog \citep{nvss} are classified as external sources.

As two of three detections have saturated optical depth profiles we set $1 - e^{-\tau} = 0.96$, following a similar approach to \citet{nickngc}, when calculating the integrated optical depth and fitting a Gaussian function later on.

\subsection{Detections}\label{sec:sources-desc}

\begin{figure*}
    \centering
\includegraphics[width=\linewidth]{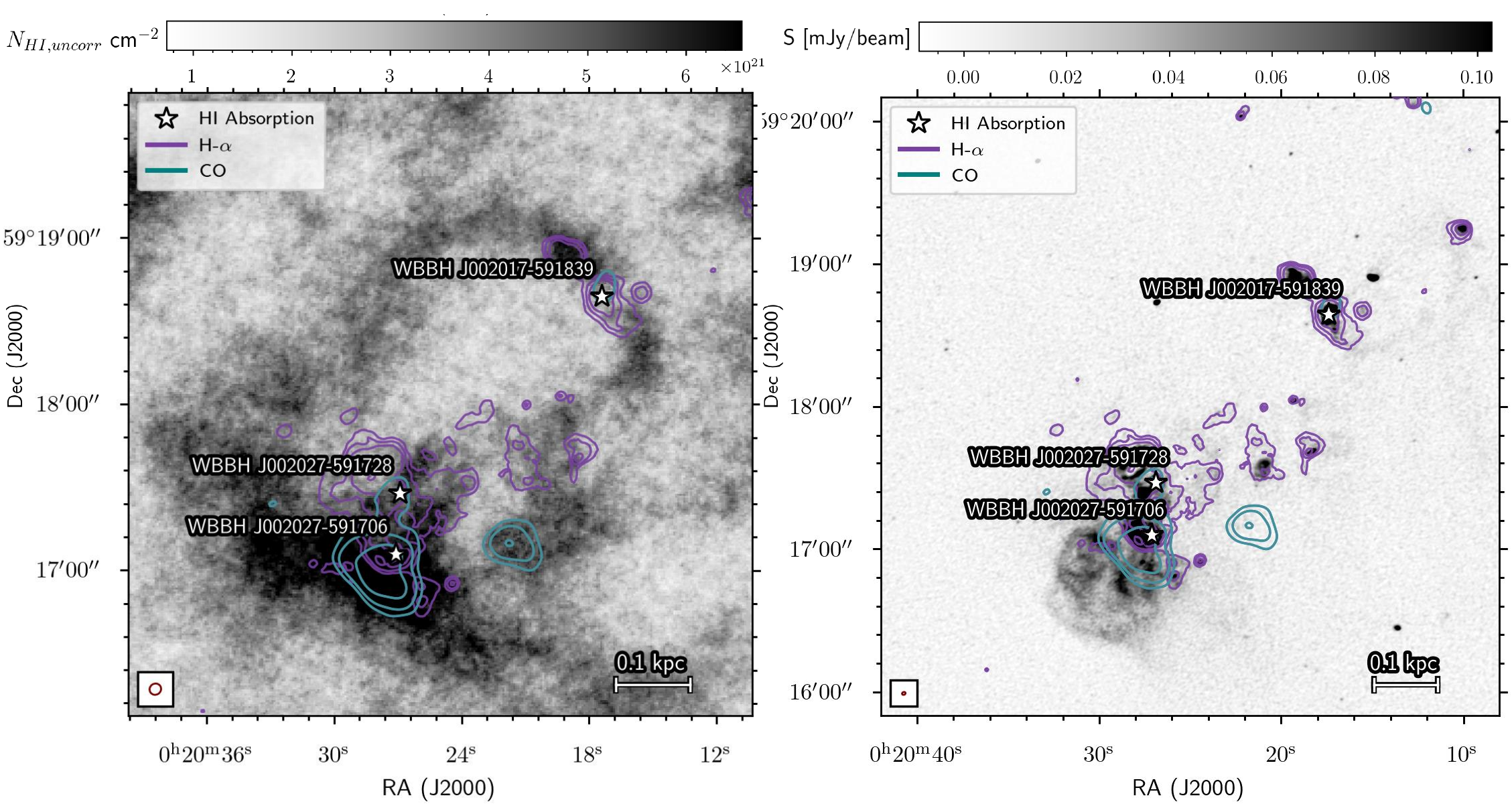}
\caption{Integrated HI intensity map (synthesized beam of $4.22'' \times 3.98''$) alongside a wide-band L-band $1.0-2.0$ GHz radio continuum image \citep[synthesized beam of $1.5'' \times 1.13''$ and rms sensitivity of $\sim 4.5 \,\,\mu\rm{Jy }\,\,\rm{ beam}^{-1}$; ][]{Corbould2025}, with the beams shown in the lower left corner of each panel. H$\alpha$ \citep{littlethings} and CO \citep{Corbould2025} contours are overlaid at $3\sigma$, $5\sigma$, and $10\sigma$ levels, shown in purple and teal, respectively. Stars indicate locations of HI absorption detections.}
\end{figure*}

We detect HI absorption toward three radio continuum sources previously identified by \citet{Westcott2017}. Two lie within $\sim1.4$–$1.7'$ of IC 10’s optical center, which we refer to as the \textit{central detections} (\textsc{WBBH J002027+591706.1} and \textsc{WBBH J002027+591728.7}), while the third, the \textit{shell detection} (\textsc{WBBH J002017+591839.7}), is located along the boundary of a shell-like structure near the edge of a large H~I hole.

The three radio continuum sources were classified as H~II regions within IC~10, based on their spectral indices \citep{Westcott2017}.
Two of these three continuum sources also exhibit modest angular extent, presented in Table \ref{tab:observations}. 
Consequently, directions toward these compact HII regions probe active, star-forming regions, and may preferentially trace denser, more turbulent CNM linked to feedback-driven structures. This contrasts with previous extragalactic absorption surveys in systems of similar metallicity like the SMC \citep[e.g.,][]{Dickey2000, Jameson2019}, which probe HI along random lines of sight mostly toward background extragalactic point-sources. 

Figure~\ref{fig:speconly} shows HI absorption spectra for the three sources (top). All three spectra have significant peak optical depth, $\tau_{\rm{peak}}>0.8$.  In the same figure (bottom) we show the corresponding HI emission spectra that are used for radiative transfer calculations in Section \ref{sec:rt,results}. The three directions probe regions of strong CO and H$\alpha$ emission, coinciding with star-forming complexes. Namely, each HI absorber is co-spatial with giant molecular clouds in IC~10 \citep[\textsc{WBBH J002027+591706.1} - B10, \textsc{WBBH J002027+591728.7} - B11, \textsc{WBBH J002017+591839.7} - B5, as discussed and identified in][]{Leroy2006}. The positions of all three sources are shown in Figure 3, overlaid with CO (teal) and H $\alpha$ (purple) intensity contours.

\textsc{WBBH J002027+591728.7} is the most compact of the three continuum sources, and its HI absorption spectrum exhibits a narrow absorption line ($\sim3$ km/s). \textsc{WBBH J002027+591706.1} and \textsc{WBBH J002017+591839.7} are slightly extended and their HI absorption profiles are very broad ($\sim13$ km/s). 
Broad absorption features may reflect enhanced turbulence or multiple cold components blended along the line of sight. A more detailed exploration of these possibilities is provided in Section~\ref{sec:discussion}.

\subsection{Non-detections}\label{sec:nondet}
Out of nine potential sightlines with sufficiently strong background sources, meeting the criteria outlined in Section~\ref{sec:obsv}, only three show detections of HI absorption, yielding a detection rate of 33\% (3/9). This is comparable to the 38\% detection rate reported by \citet{Dempsey} for the SMC. Their statistics are based on a much larger sample of 162 sightlines that were evenly distributed across regions with column densities above $6 \times 10^{20}\,\mathrm{cm}^{-2}$, a threshold chosen by the authors to enclose most of the SMC’s HI mass.

\begin{figure}
    \centering
    \includegraphics[width=\linewidth]{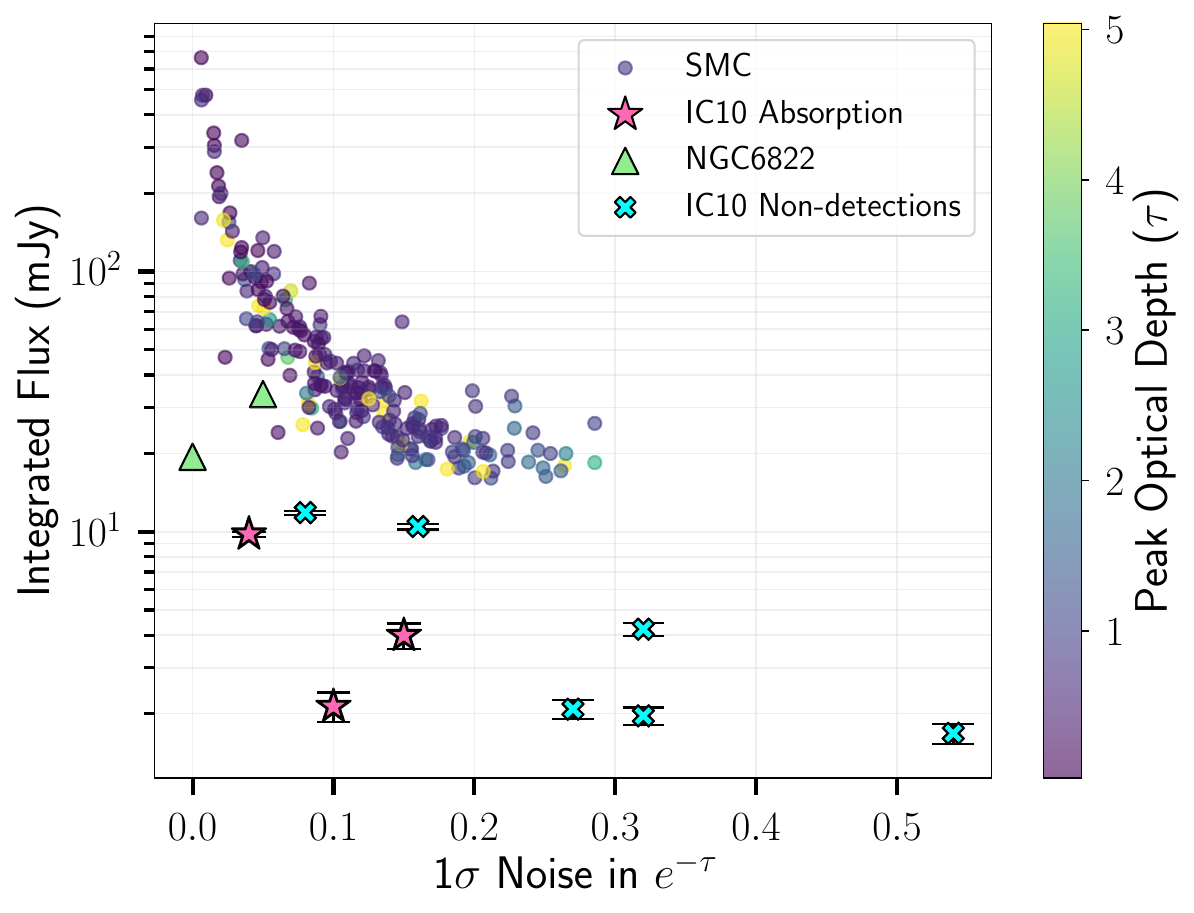}
    \caption{Peak continuum flux density versus $1\sigma$ noise in $e^{-\tau}$ for absorption detections in IC~10 (pink stars), the SMC from \citet{Dempsey} (circles, color-coded by peak optical depth $\tau$), and NGC 6822 (green triangles). IC~10 and NGC 6822 absorption spectra were smoothed using a Gaussian kernel to achieve similar spectral resolution to GASKAP. The peak flux density of the LGLBS background sources has been scaled to address the difference in beam widths between the two surveys. IC~10 points lie systematically below the SMC, as the surveyed region contains fewer bright sources. Fluxes and $1\sigma$ noise in $e^{-\tau}$ are scaled to the GASKAP beam and spectral resolution for a fair comparison.
}
    \label{fig:flxvsnoise}
\end{figure}

To compare the sensitivity achieved by LGLBS for IC~10 with that of the SMC, we first smooth our IC~10 absorption spectra to match the 1 km s$^{-1}$ velocity resolution used in the SMC study. Figure~\ref{fig:flxvsnoise} shows the $1\sigma$ rms noise in the absorption spectra ($e^{-\tau}$) as a function of the integrated flux of the background source. This figure demonstrates that, for our detections, the optical depth sensitivity is comparable to that of the SMC survey. However, in 3 out of 6 non-detections, the noise levels are higher than those achieved in the SMC by \citet{Dempsey}, limiting our ability to detect weak absorption features ($\tau<0.1$).

The continuum sources in IC~10 are also intrinsically fainter than those in the GASKAP SMC survey, with integrated fluxes typically below 15 mJy, compared to 20–100 mJy for most SMC sources (see Table~\ref{tab:observations}). 
While the SMC and IC~10 have comparable physical sizes (radii of $\sim$2.5–3 kpc), IC~10's greater distance results in a 0.25 deg$^2$ solid angle area spanned on the sky, which is two orders of magnitude smaller than the $\sim$25 deg$^2$ area covered by the SMC. This significantly limits the number of bright background radio continuum sources available for HI absorption studies toward IC~10. 
As a result, our observations are mostly sensitive to absorption against the few relatively strong continuum sources \textit{within} the gas-rich parts of the galaxy. 

Most non-detections correspond to spectra with rms noise in $e^{-\tau}$ greater than 0.25, 
and are probing regions outside the central region of IC~10, with a median HI column density of $N \approx 5.1 \times 10^{20}$ cm$^{-2}$. This is nearly an order of magnitude lower than the typical column densities in regions where absorption is detected ($N \gtrsim 1.57 \times 10^{21}$ cm$^{-2}$), highlighting that the non-detections probe lower-density environments.

Future, more sensitive observations will be necessary to probe a comparable population of CNM clouds as seen in the SMC. Specifically, a $3\sigma$ detection of $\tau = 0.1$ would require sightlines with $\sigma_{e^{-\tau}} \lesssim 0.12$. In the case of NGC~6822 which was observed as part of LGLBS \citep{Pingel2024}, a similar sensitivity to the present study was achieved in the emission line cubelets. However, the two HI absorption detections were measured against radio continuum background sources with integrated fluxes an order of magnitude higher than those available for IC~10 (19.4 and 33.8 mJy).

\section{Analysis}\label{sec:rt} 
\subsection{Gaussian Decomposition and Radiative Transfer Calculations}
We perform Gaussian decomposition of the absorption and corresponding emission spectra to derive physical properties of individual HI structures along each lines of sight. The method we are employing follows that originally outlined in \cite{Heiles_2003} (hereafter referred to as \citetalias{Heiles_2003}), with additional optimizations and best-fit testing introduced by \cite{Chen2025a}. 

Under a two-phase HI framework \citep{Dickey2003}, \citetalias{Heiles_2003} assumes that the observed HI brightness temperature is the sum of contributions from both the CNM and WNM:

\begin{equation}\label{eq:total}
    T_{\mathrm{B,\,obs}}(v) = T_{\mathrm{B,\,CNM}}(v) + T_{\mathrm{B,\,WNM}}(v),
\end{equation}
\noindent
where $T_{\mathrm{B,\,CNM}}(v)$ represents the brightness temperature due to gas detected in absorption, arising from the CNM, and $T_{\mathrm{B,\,WNM}}(v)$ denotes the contribution from WNM gas, which is not seen in absorption and appears only in emission.

We begin by modeling the optical depth $\tau(v)$ from the observed HI absorption spectrum. This is achieved by fitting an appropriate number of $N$ Gaussian functions,  each representing a distinct CNM structure along the line of sight. The total sum of the $N$ components makes up the total optical depth contribution:

\begin{equation}\label{eq:tau}
     \tau(v) = \sum_{n=0}^{N-1} \tau_{0,n} \times e^{-4\ln2[(v-v_{0,n})/\Delta v_n]^2}
\end{equation}
where $\tau_{0,n}$ is the peak optical depth, $v_{0,n}$ is the central velocity and  $\Delta v_n$ is the FWHM of component $n$.
Assuming that each fitted component is independent and isothermal, characterized by a spin temperature $T_{\rm{s, n}}$, the corresponding contribution of these $N$ Gaussian CNM components to the expected brightness temperature is expressed as:

\begin{equation}\label{eq:tbcnm}
    T_{B, \text{CNM}}( v) = \sum_{n=0}^{N-1} T_{s,n}(1-e^{-\tau_n(v)}) e^{-\Sigma^{M-1}_0 \tau_m(v)}
\end{equation}
Here, $\tau_m(v)$ denotes the optical depth of each of the $m$ CNM clouds located in front of cloud $n$, and the exponential term accounts for their cumulative attenuation.  

In modeling the remaining contribution to the expected brightness temperature profile, we use $k$ Gaussians to represent the unabsorbed emission from the WNM:
\begin{align}\label{eq:tbwnm}
    T_{B, \mathrm{WNM}}(v) &= \sum_0^{K-1}\bigg[F_k+\left(1-F_k\right) e^{-\tau(v)}\bigg] \times \nonumber \\
&\quad T_{0, k} e^{-4 \ln 2 \left[\left(v-v_{0, k}\right) / \Delta v_k\right]^2}.
\end{align}
Here, $T_{0,k}$, $v_{0,k}$, and $\Delta v_k$ are the Gaussian fitting parameters representing the peak brightness temperature, central velocity, and FWHM of the unabsorbed $k$-th WNM component. The parameter $F_k$  characterizes the amount of WNM emission that remains unattenuated, reflecting the fraction of WNM lying in front of the CNM. Given a specified set of absorption components, the expected profile in Equation~\ref{eq:total} is fit by simultaneously solving for the spin temperatures of the $N$ CNM components and the Gaussian parameters of the $K$ WNM emission components. We set a minimum threshold on $T_{0,k}$ equal to the average 1-$\sigma$ noise level of the emission spectra to prevent spurious low-amplitude components. 

In our radiative transfer calculations, we adopt a fixed value of $F_k = 0.5$ for all WNM components. This choice is physically motivated by the geometry of the system: the continuum sources with detected HI absorption are compact HII regions embedded within IC~10, rather than background point sources. As such, the detection of HI absorption requires that the absorbing CNM lies in front of the continuum source, while a substantial portion of the neutral medium must reside behind it to account for the galaxy’s full line of sight extent. Fixing $F_k$ to 0.5 provides a reasonable approximation of this configuration, reduces model degeneracy, and helps constrain the spin temperature solutions ($T_{\rm s, n}$), decreasing the uncertainty value than if more variation in the WNM configuration were allowed.

Given this geometry, any CNM located behind the continuum source does not contribute to the absorption but still contributes to the emission. Likewise, the assumed fraction of WNM lying in front of the CNM ($F_k = 0.5$) affects the absorption/emission modeling. As a result, the derived CNM column densities and cold gas fractions ($f_{\rm CNM}$) should be interpreted as lower limits, since the clouds behind the continuum or the WNM effectively treated as foreground are counted as WNM in the model, and the true CNM content could be higher.

The final best-fit model for each sightline is determined by minimizing the residuals between the observed and modeled spectra (see Section~4.1 of \cite{Chen2025a}). The corresponding spin temperature $T_{\rm{s, n}}$ is then computed as a weighted average across all fitting trials, following Equations~(21a) and (21b) of \citetalias{Heiles_2003}.
\begin{table*}[ht!]
\centering
\small 
\renewcommand{\arraystretch}{1.0}
\setlength{\tabcolsep}{2pt} 
\begin{tabular}{ccccccccc}
\hline
Source Name & $f_{\text{CNM}}$ & Phase & $T_s$ (K) & $\tau_{\text{peak}} \,\text{or}\,\,T_B$ (K) & $v_0$ (km s$^{-1}$) & $\Delta v_0$ (km s$^{-1}$) & $T_{\rm{k, max}}$ (K) & $N_{\text{HI}}$ ($10^{21} \text{cm}^{-2}$) \\
\hline
\textsc{WBBH J002027+591706.1} & 0.37 $\pm$ 0.06 & CNM & 54.7 $\pm$ 3.3 & 2.12 $\pm$ 0.10 & -328.66 $\pm$ 0.28 & 12.44 $\pm$ 0.66 & 3389 & 2.80 $\pm$  0.26\\
&  & WNM1 &  & $68.18 \pm 0.74$ & -327.67 $\pm$ 3.32 &  33.72 $\pm$ 8.78  & 24863 & 4.46$\pm$ 1.16 \\
& & WNM2 & & 21.82 $\pm$ 0.39 & -368.60 $\pm$ 0.48 & 7.69 $\pm$ 5.88 & 1295 & 0.32 $\pm$ 0.2  \\ 
\hline      
\textsc{WBBH J002027+591728.7} & 0.21 $\pm$ 0.04  & CNM & 37.7 $\pm$ 3.7 & 2.03 $\pm$ 0.10 & -332.50 $\pm$ 0.14 & 5.58 $\pm$ 0.39 & 681 & 0.83 $\pm$ 0.1 \\
&  & WNM1 &  & 77.10 $\pm$ 2.74  & -334.80 $\pm$ 3.71 & 19.23 $\pm$ 8.95 & 8090 & 0.28 $\pm$ 0.13  \\
&  & WNM2 &  & 23.59 $\pm$ 0.28 & -354.20 $\pm$ 0.30 & 4.00 $\pm$ 8.14 & 349 & 0.18 $\pm$ 0.37 \\
\hline
\textsc{WBBH J002017+591839.7} & 0.32 $\pm$ 0.17 & CNM &  30.6 $\pm$ 10.6 & 0.83 $\pm$ 0.03   & -341.42 $\pm$ 0.22 & 13.58 $\pm$ 0.54 & 4032 & 0.67 $\pm$ 0.23   \\
& & WNM1 &  & 27.00 $\pm$ 2.88  & -329.20 $\pm$ 10.6 & 24.65 $\pm$ 17.45  &  13284 &  1.29 $\pm$ 0.92 \\
& & WNM2 & & 15.96 $\pm$ 1.40  & -353.30 $\pm$ 1.50   & 3.85 $\pm$ 8.05  &  324 & 0.12 $\pm$ 0.2 \\
\hline

\end{tabular}
\caption{Radiative transfer results for CNM and WNM in IC~10, obtained via Gaussian decomposition.}
\label{tab:radtrans}
\end{table*}

\section{Physical properties of CNM in IC~10}\label{sec:results}
\begin{figure*}
    \centering
\includegraphics[width=\linewidth]{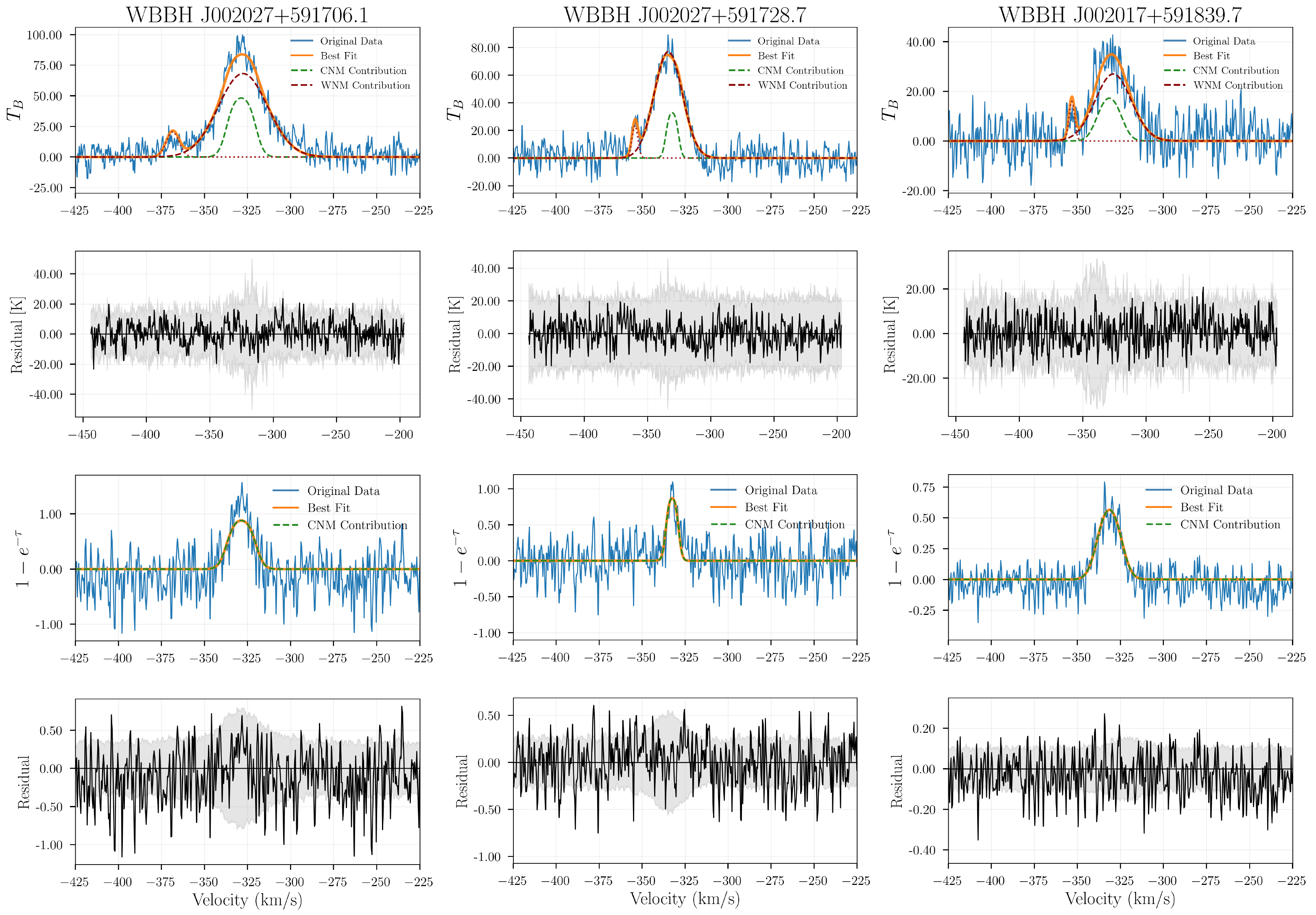}
\caption{Emission and absorption spectra for the three sources, along with their Gaussian decompositions and residuals. The top panels display the HI emission spectra, while the bottom panels show the optical depth spectra, plotted as $(1 - e^{-\tau})$. The original spectra are shown in blue, with the best-fit model from radiative transfer calculations (see Section \ref{fig:radtrans}) overlaid in orange. In the emission panels, the contribution of each WNM component is plotted in red, and the contribution of each CNM component to the emission is shown in green. In the absorption panels, the contribution of each individual CNM component is shown in green.The fitting parameters are provided in Table~\ref{tab:radtrans}}
    \label{fig:radtrans}
\end{figure*}
\subsection{Fitting results}\label{sec:rt,results}
Figure \ref{fig:radtrans} presents the HI emission and absorption spectra for the three sources, along with their phase decomposition, highlighting the contributions of the WNM and CNM components. The corresponding Gaussian fitting parameters are shown in Table \ref{tab:radtrans}, where we present the source name, the fraction of cold gas ($f_{\text{CNM}}$), the gas phase (CNM or WNM), the spin temperature ($T_s$), the peak optical depth ($\tau_{\text{peak}}$) for the CNM components or brightness temperature ($T_B$) for the WNM components, the peak central velocity ($v_0$), the full width at half maximum (FWHM, $\Delta v_{0}$), the maximum kinetic temperature ($T_{k,\text{max}}$), and the neutral hydrogen column density of each phase component ($N_{\text{HI}}$). 

As described in Section~\ref{sec:sources-desc}, we refer to the two sources near IC~10’s optical center as the \textit{central detections} (\textsc{WBBH J002027+591706.1} and \textsc{WBBH J002027+591728.7}), and to the third, located along the boundary of a shell-like structure, as the \textit{shell detection} (\textsc{WBBH J002017+591839.7}).

The central detections exhibit saturated absorption spectra. As previously discussed, for these considerably saturated spectra we set a maximum threshold of $1 - e^{-\tau} = 0.96$ to avoid infinite values when converting the absorption to optical depth for fitting purposes. After applying this threshold, we measure peak optical depths of $2.12$ and $2.03$, which should be interpreted as lower limits. In such cases, the standard optically thin approximation (Equation~\ref{eq:nhi,thin}) underestimates the true HI column density. For the shell detection, the absorption spectrum is not saturated, and we measure a peak optical depth of $0.83$.

For each of 3 sources, the total corrected column density ($N_{\text{HI}}$), representing the sum of all the WNM and CNM components along the line of sight, is: $7.5 \times 10^{21}$ cm$^{-2}$ for \textsc{WBBH J002027+591706.1}, $3.8 \times 10^{21}$ cm$^{-2}$ for \textsc{WBBH J002027+591728.7}, and $2.1 \times 10^{21}$ cm$^{-2}$ for \textsc{WBBH J002017+591839.7}. 

We can then estimate a lower limit on the correction factor, $\mathscr{R}_{\mathrm{HI}}$, defined as the ratio of $N_{\mathrm{HI}}$ and $N_{\mathrm{HI,uncorr}}$. As reported in Table \ref{tab:observations}, the uncorrected column densities are $5.6 \times 10^{21}$ cm$^{-2}$ for \textsc{WBBH J002027+591706.1}, $3.0 \times 10^{21}$ cm$^{-2}$ for \textsc{WBBH J002027+591728.7}, and $1.5 \times 10^{21}$ cm$^{-2}$ for \textsc{WBBH J002017+591839.7}. For the three sources, we find $\mathscr{R}_{\mathrm{HI}} = 1.34$, $1.28$, and $1.32$, respectively. This suggests that the bright central star-forming regions of IC~10 have the HI mass underestimated by at least 20-40\% if the optically thin assumption is applied.

We report similar central velocities for sources along similar RA and DEC: absorption in the central detections peaks at  $-328.66$ km s$^{-1}$ and $-332.50$ km s$^{-1}$ respectively, whereas the shell detection has a central velocity for the HI absorption component of $-341.42$ km s$^{-1}$.

The resulting spin temperatures of the three CNM components we have identified are in the range $30-60$ K. The lowest spin temperature is observed for \textsc{WBBH J002017+591839.7} ($30.6 \pm 10.6$ K), while \textsc{WBBH J002027+591706.1} exhibits the highest ($54.7 \pm 3.3$ K). \textsc{WBBH J002027+591728.7} shows an intermediate CNM temperature of $37.7 \pm 3.7$ K. 

The kinetic temperature, $T_{\rm k, max}$, is calculated from the observed line widths under the assumption that the line broadening is purely thermal. For a completely thermalized gas, the maximum kinetic temperature is given by:
\begin{equation}
    T_{\rm k, max} = \frac{m_{\rm H}}{8 k_{\rm B}\ln 2}\Delta v_0^2 =21.866 \times\Delta v_0^2
\end{equation}
where $m_{\rm H}$ is the hydrogen atom mass, $k_{\rm B}$ is the Boltzmann constant, and $\Delta v_0$ is the FWHM of the line \citep{Draine2011}. We estimate $T_{\rm{k, max}} = 3389$ K for \textsc{WBBH J002027+591706.1}, $T_{\rm{k, max}} = 681$ K for \textsc{WBBH J002027+591728.7}, and $T_{\rm{k, max}}= 4032$ K for for \textsc{WBBH J002017+591839.7}.

In terms of linewidth, for two identified CNM components, we measure broad line widths ($> 12$ km s$^{-1}$) - \textsc{WBBH J002027+591706.1} has a FWHM of $12.44 \pm 0.66$ km s$^{-1}$ and \textsc{WBBH J002017+591839.7} has a FWHM of $13.58 \pm 0.12$ km s$^{-1}$. These profiles were obtained against slightly extended background sources (see Table ~\ref{tab:observations} for information on their angular extent), which might contribute to the line broadening. In contrast, \textsc{WBBH J002027+591728.7}, extracted against a more compact source, exhibits the narrowest HI absorption line of the three sightlines of $5.58 \pm 0.39$ kms$^{-1}$.

All three sources exhibit two distinct WNM components that follow a consistent pattern: a broad WNM component with a corresponding peak central velocity closely aligned with the CNM component, and a second, narrower WNM component whose line width is comparable to or even smaller thant that of the CNM in emission. Specifically, \textsc{WBBH J002027+591706.1} shows broad and narrow WNM components with linewidths of $33.72 \pm 8.78$ and $7.69 \pm 5.88$ km s$^{-1}$ at $v_{\rm peak} = -327.67$ and $-368.60$ km s$^{-1}$, respectively. \textsc{WBBH J002027+591728.7} exhibits components of $19.23 \pm 8.95$ and $4.00 \pm 8.14$ km s$^{-1}$ at $v_{\rm peak} = -334.80$ and $-354$ km s$^{-1}$, with the latter narrower than the CNM. Lastly, \textsc{WBBH J002017+591839.7} shows linewidths of $24.65 \pm 17.45$ and $3.85 \pm 8.05$ km s$^{-1}$ at $v_{\rm peak} = -329$ and $-353$ km s$^{-1}$. The coexistence of broad and narrow WNM components suggests the presence of multiple thermal phases along the same line of sight, with the broad component tracing the classical, thermally stable WNM, while the narrow component may represent thermally unstable neutral medium (UNM) or unresolved CNM substructure within the emission beam.

%%% FCNM
By directly probing the CNM through HI absorption, we are able to estimate the cold gas fraction, $f_{\text{CNM}}$. Given the derived spin temperature $T_s$, we compute the CNM contribution to the total HI column density along each line of sight with detected absorption. The total CNM and WNM column densities for each line of sight are obtained by summing over all individual components, denoted as $N_{\text{HI, CNM, all}}$ and $N_{\text{HI, WNM, all}}$, respectively. The cold gas fraction is defined as:
\begin{equation}\label{eq:fcnm}
    f_{\text{CNM}} =
    \frac{  N_{\text{HI, CNM, all}}}
    {N_{\text{HI, CNM, all}} + N_{\text{HI, WNM, all}}}
\end{equation}

For the three sightlines in IC~10 with detected absorption, we estimate $f_{\text{CNM}} = \{0.37, 0.21, 0.32\}$. These values should be considered lower limits, as our analysis is limited to continuum sources embedded within IC~10 and two out of the three sightlines suffer from additional saturation effects. Any cold HI located behind these bright HII regions remains undetectable in absorption, and thus does not contribute to our estimate of $N_{\text{HI, CNM}}$. Consequently, the true cold gas fraction along these lines of sight may be higher than what we report here.
\subsection{Linewidths of HI Absorption and Molecular Line Emission}\label{sec:linewidths}
\begin{figure*}
    \centering
    \includegraphics[width=\linewidth]{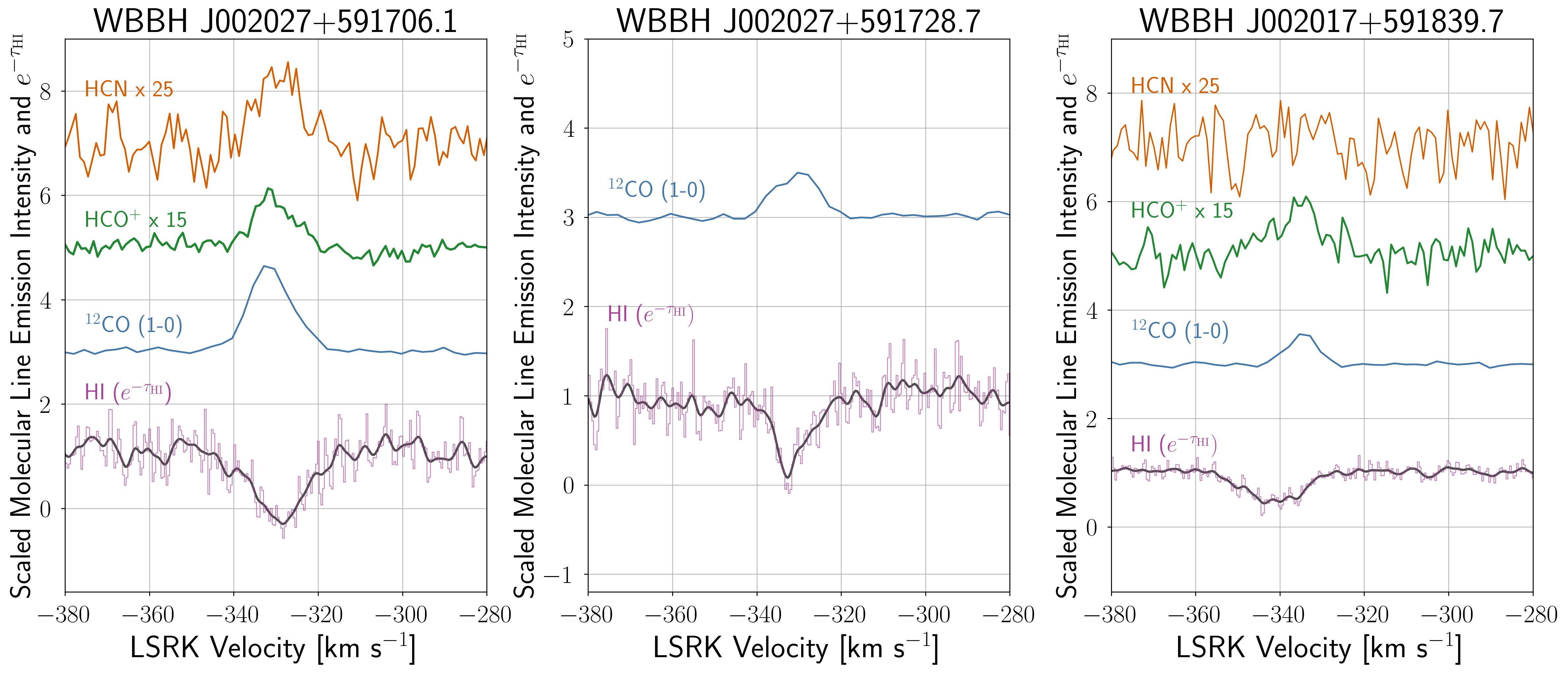}
    \caption{HI absorption and molecular line spectra toward WBBH J002027+591706.1 (left), WBBH J002027+591728.7 (middle), and WBBH J002017+591839.7 (right). In each panel, the darker purple curves show smoothed HI absorption spectra, plotted as $e^{-\tau_{\rm HI}}$ and scaled for clarity; the HI absorption was smoothed with a Gaussian kernel matched to the spectral resolution of the molecular data (5 km s$^{-1}$). Colored solid lines indicate emission from $^{12}$CO (1–0) (blue), HCO$^+$ (green), and HCN (orange) \citep{kepley2018}, each vertically offset to emphasize spectral features. The relative velocity alignment of the gas components highlight a tightly coupled atomic and molecular medium.}
    \label{fig:molecules}
\end{figure*}

We compare the HI radiative transfer solutions to other cold phases of the interstellar medium, focusing in particular on the cold phase traced by molecular line emission. Figure \ref{fig:molecules} shows the HCO$^+$,  HCN and $^{12}$CO(1-0) emission spectra from \cite{kepley2018} (green, orange and blue respectively), alongside the HI absorption spectra from this work, smoothed to match the 5 km s$^{-1}$ velocity resolution of the HCO$^+$ and HCN data. Measurements of the HCO$^+$ and HCN lines were obtained using the GBT, which has a resolution of $8.5''$ corresponding to  $\sim \,34$ pc, approximately the size of an individual giant molecular cloud. $^{12}$CO(1-0) data, hereinafter referred to as CO, is combined from observations with CARMA and the IRAM 30m telescope, yielding a synthesized beam of $\sim , 8.5''$ and a velocity resolution of 2.5 km s$^{-1}$ (for more information on these observations, we refer the reader to \cite{kepley2018}.)

Figure \ref{fig:molecules} shows that the HI absorption and molecular line emission are found at similar radial velocities and all spectral features have relatively similar line widths. This suggests that the molecular gas is well-mixed with the HI. 

In the diffuse molecular regions in the Milky Way, \cite{dan2022} found that HI absorption profiles are generally much broader than those of molecular species such as HCO$^+$, HCN, and HNC, with only their broad and low optical depth HCO$^+$ components approaching the width of HI. In contrast, our observations in IC~10 show that the CO, HCO$^+$, and HCN emission features exhibit linewidths relatively comparable to those of the HI absorption. This suggests that the CNM in IC~10 is denser than the CNM in the Milky Way, as expected for low metallicity conditions.

\textsc{WBBH J002027+591706.1} exhibits an HI absorption feature with a FWHM of $13.60 \pm 0.78$ km s$^{-1}$ and a central velocity of $-328.93 \pm 0.03$ km s$^{-1}$ from Gaussian fitting. This line of sight intersects GMC B11 (as seen in \cite{Leroy2006}) and HII region [HL106]. The CO emission has a FWHM = $11.2 \pm 0.07$ km s$^{-1}$ and a central velocity of $v_\mathrm{CO} = -331.6$ km s$^{-1}$. A young massive cluster (YMC) candidate of stellar mass $M_\star \approx (8.2 \pm 1.0) \times 10^3\,M_\odot$ is also present at this exact location \citep{Bouchereau}. As reported by \citet{kepley2018}, the HCO$^+$ line has a FWHM of $11.0 \pm 0.6$ km s$^{-1}$ at $v_{\rm c} = -330.8 \pm 0.3$ km s$^{-1}$, while the HCN line has a FWHM of $11.2 \pm 0.3$ km s$^{-1}$ at $v_{\rm c} = -329.4 \pm 0.6$ km s$^{-1}$.

\textsc{WBBH J002027+591728.7} exhibits the narrowest HI absorption line (FWHM of $5.58 \pm 0.39$ kms$^{-1}$) and a central velocity of $-332.5$ km s$^{-1}$. It is associated with GMC B10, ($v_\mathrm{CO} = -330.5$ km s$^{-1}$, FWHM = 12.06 km s$^{-1}$), HII region [HL 111] and a YMC with $M_\star \approx 1.6 \pm 0.7 \times 10^4$ M$_\odot$ \citep{Bouchereau}. There were no pointings towards this sightline in the \citeyear{kepley2018} study.

\textsc{WBBH J002017+591839.7} lies along the rim of the expanding HI hole \citetalias{Wilcots}7 with an expansion velocity of $\sim 16$ km s$^{-1}$ \citep{holes2020}. The HI absorption has a FWHM of $13.58 \pm 0.12$ km s$^{-1}$ and a central velocity of $-341.1$ km s$^{-1}$. This source coincides with GMC B5 ($v_\mathrm{CO} = -335.3 $ km s$^{-1}$, FWHM = 8.4 km s$^{-1}$), HII region [HL 45] and a YMC of mass $M_{\star} \approx \, 2.3 \pm 0.6\times 10^4$ M$_{\odot}$ estimated from 225 GHz flux of ($6.6 \pm 1.8$) mJy \citep{Bouchereau}. At this sightline, only HCO$^+$ was found in emission, with a FWHM of $14.8 \pm 1.7$ km s$^{-1}$ at $v_{\rm c} = -335.6 \pm 0.7$ km s$^{-1}$.

We measure the velocity offset between CO and HI, $\Delta v =|v_{\rm{CO}}-v_{\rm{HI}}|$ for each individual sightline in IC~10. \textsc{WBBH J002027+591706.1} and \textsc{WBBH J002027+591706.1} exhibit relatively small offsets, of 2.67 and 2 km s$^{-1}$ respectively, whereas \textsc{WBBH J002017+591839.7} shows a slightly larger offset of 5.8 km s$^{-1}$ signaling a more dynamic environment.
The HCO$^+$ line profile in the direction of \textsc{WBBH J002017+591839.7} exhibits an unusually large FWHM of 14.8 km s$^{-1}$ and elevated line intensities \citep{kepley2018}. These features were difficult to explain solely as a consequence of line blending within the GBT’s $8''$ beam. \cite{kepley2018} attributed the enhanced HCO$^+$ emission to shock-driven processes, pointing out that the sightline intersects several expanding bubbles (see Figure~1 in \citealt{kepley2018}; see also \citealt{Wilcots,Leroy2006}).

\section{Discussion}\label{sec:discussion}
Until LGLBS, the only high-resolution views of cold HI in low-metallicity galaxies, directly detected via HI absorption, were provided by the Magellanic Clouds \citep{Dempsey, Chen2025a, Chen2025b}. In particular, the SMC with its gas-phase metallicity of $\sim0.25\,Z_\odot$ has served as an example of a primitive ISM environment similar to galaxies at high redshift. The high-resolution and sensitivity observations with LGLBS provided direct detections of cold HI in absorption in NGC 6822 \citep{nickngc} and IC~10.  

% Due to the solid angle occupied by NGC 6822 and IC~10 being much smaller than that of the SMC and LMC, the sample of HI absorption measurements from LGLBS is limited by the small number of bright continuum sources where our sensitivity is sufficient to detect cold HI.

NGC 6822 and IC10 cover a significantly smaller area on the sky compared to the Magellanic Clouds ($\sim0.25$ deg$^2$ vs. $\sim25$ deg$^2$ respectively).
% \textbf{The linear resolution achieved in NGC6822 and IC10 is significantly coarser than in the SMC and LMC. 
As a result,  the sample of HI absorption measurements from LGLBS is limited by the small number of bright continuum sources where our sensitivity is sufficient to detect cold HI. In fact, while surveys of HI absorption in the SMC largely sample background (point) sources, our IC~10 absorption detections are towards solely the brightest HII regions in IC~10. With only three HI absorption spectra it is not possible to undertake a statistical analysis of cold HI properties; however, we can provide a glimpse at key CNM characteristics and compare CNM in IC~10 with what is known about CNM properties in other low-metallicity galaxies.

\subsection{Integrated Properties of Cold HI in IC10 and other dwarf galaxies}
\begin{figure}
    \centering
    \includegraphics[width=\linewidth]{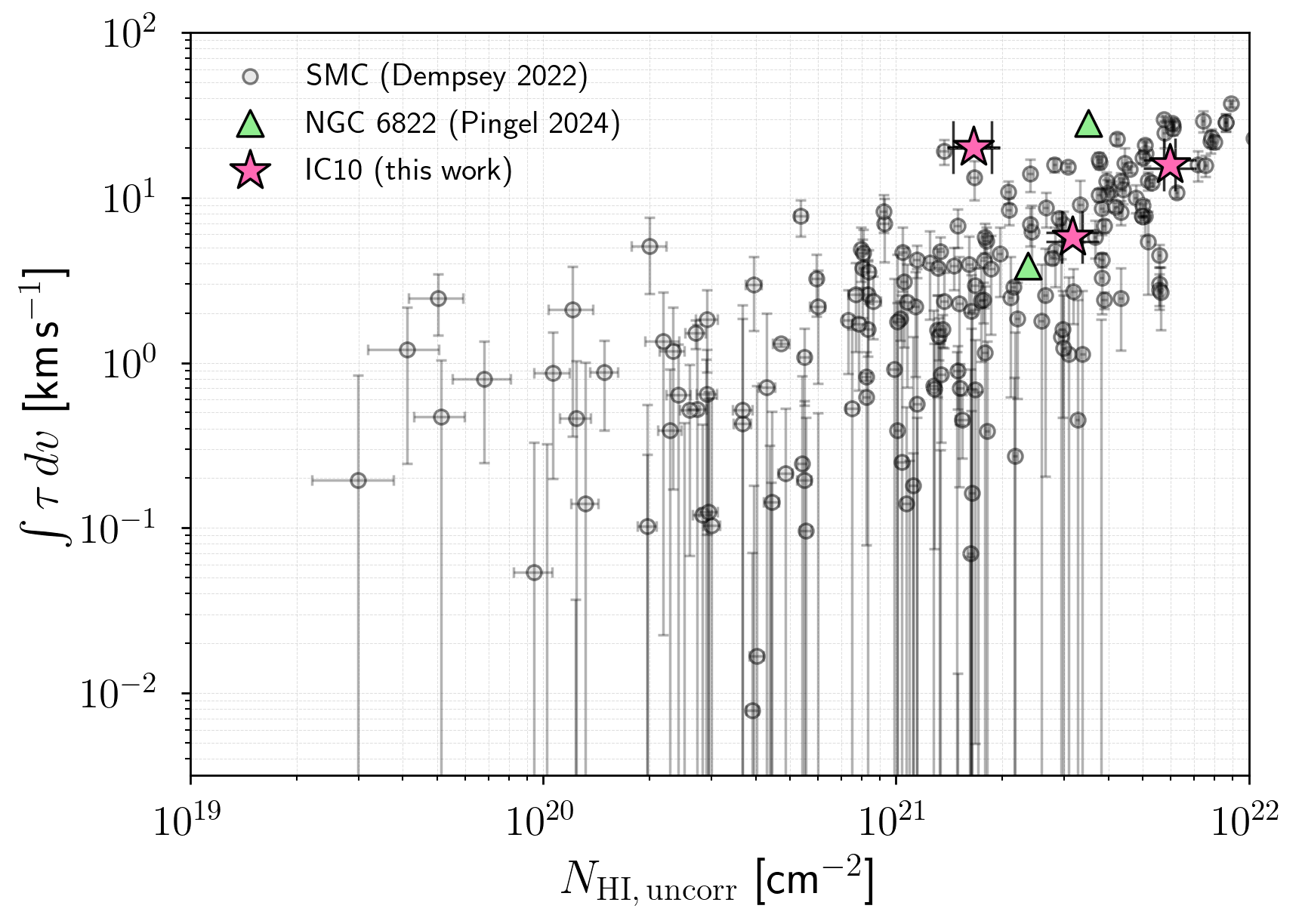}
    \caption{Uncorrected H~I column density ($N_{\mathrm{HI,uncorr}}$), derived under the optically thin assumption (Eqn.~\ref{eq:nhi,thin}), versus integrated optical depth ($\int \tau\, dv$) for IC~10 (pink stars; this work). For comparison, measurements from the SMC \citep[light gray circles;][]{Dempsey} and NGC~6822 \citep[green triangles;][]{nickngc} are shown. Both axes are logarithmic; error bars indicate $1\sigma$. Negative SMC values are excluded.}
    \label{fig:tauvsnhi}
\end{figure}

Figure~\ref{fig:tauvsnhi} shows the integrated optical depth ($\int \tau \, dv$) as a function of uncorrected column density, calculated under the optically thin assumption ($N_{\rm HI,uncorr}$; Eqn.~\ref{eq:nhi,thin}) for our three detections. To place the level of HI absorption detected in IC~10 in context with other galaxies of similar metallicity, measurements of the same physical quantities from the SMC \citep{Dempsey} and NGC~6822 \citep{nickngc} are included. 

In terms of integrated properties, the CNM detected in IC~10 traces the high column density ($N_{\rm HI} > 10^{21}$ cm$^{-2}$) and high optical depth ($\tau > 0.8$) end of the CNM population observed in the SMC, resembling that found in NGC~6822. As shown in Figure~\ref{fig:tauvsnhi}, the IC~10 detections clearly probe the high $\int \tau \, dv/N_{\mathrm{HI}}$ regime of the CNM population, given our current sensitivity. This suggests that future, deeper observations could potentially reveal CNM components with lower optical depths and column densities. 

\subsection{Individual CNM Components in IC10 and other dwarf galaxies}
Similarly to our previous comparison, we next compare the radiative transfer–derived cold gas properties of IC~10 with those of dwarf galaxies of similar metallicity, namely NGC~6822 \citep[observed as part of LGLBS;][]{nickngc}, and the outskirts of the SMC and LMC \citep[observed as part of GASKAP-HI;][]{Dempsey, Chen2025a}. For the following, also include measurements of cold gas in the main body of the LMC \citep{Chen2025b}.

\subsubsection{Spin Temperature}
We measure spin temperatures of 30–60 K for our three detections, with a mean value of 44 K. These values are comparable to those of the five CNM components identified in NGC~6822, which have a mean $T_{\mathrm{s}}$ of 32 K, and to those found in the outskirts of the SMC \citep[median 27 K;][]{Chen2025a}. In the outskirts of the LMC, CNM components exhibit a similar median value of 24 K \citep{Chen2025a}, while in its main body, the spin temperature is higher, with a median of $\sim$35 K and a mean of $\sim$50 K \citep{Chen2025b}. Across all dwarf galaxies the spin temperatures are broadly consistent around $\sim$25–60 K. For comparison, the Milky Way exhibits a median CNM spin temperature of $\sim$50 K and a broader CNM temperature distribution spanning $\sim$50–200 K \citep{naomi}. This implies that the CNM in dwarf galaxies is systematically cooler by roughly a factor of $1.5–4$ relative to the upper end of the Milky Way CNM distribution, and remains confined to the lower end of the Galactic temperature range.

% , converging to values systematically lower than those found in the Milky Way.}

\subsubsection{Line widths and turbulent broadening}\label{sec:line}
\begin{figure}
\centering
\includegraphics[width=\linewidth]{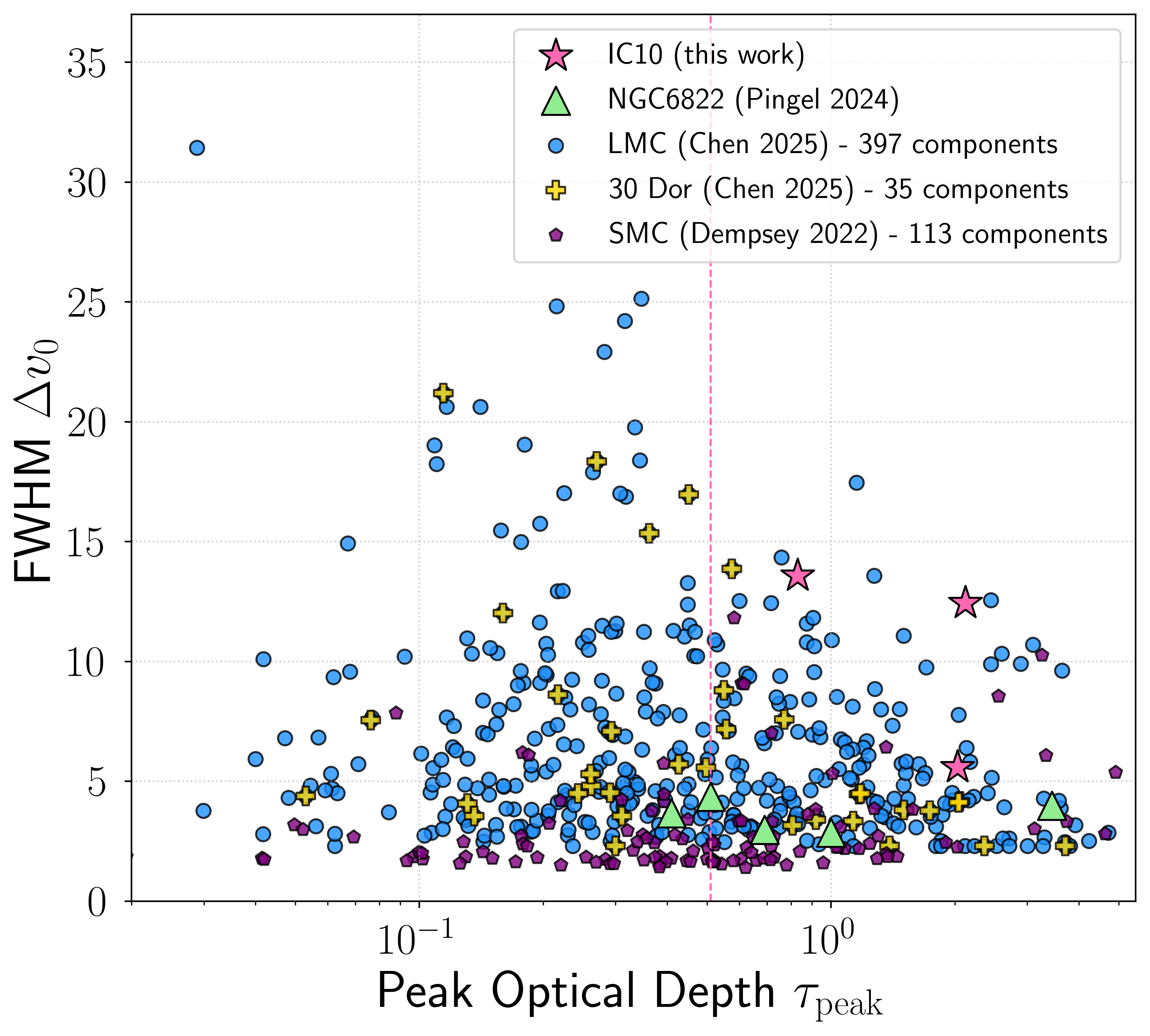}
    \caption{\textit{Comparison of CNM components properties across Local Group galaxies.} Peak optical depth ($\tau_{\rm{peak}}$) versus FWHM line width ($\Delta v_0$) for HI absorption components detected toward background sources in IC~10 (this work; pink stars), NGC 6822 \citep[green triangles;][]{Pingel2024}, the SMC \citep[purple pentagons;][]{Dempsey}, and the LMC \citep[blue circles;][]{Chen2025b}. For the LMC components, sightlines within 30 arcmins away from 30~Dor complex are marked by the gold crosses. A vertical dashed line marks the $3\sigma$ optical depth sensitivity threshold of this study ($\tau_{\rm{peak}} = 0.51$). Sources with broader components ($\Delta v_0 > 10$ km s$^{-1}$) are present in each sample, potentially indicating line blending or turbulent gas structures.}
    \label{fig:gp}
\end{figure}
Figure \ref{fig:gp} shows the peak optical depth, $\tau_{\rm{peak}}$, and FWHM $\Delta v_0$ for the individual Gaussian components of absorption profiles observed in NGC~6822 \citep[green triangles;][]{nickngc}, the LMC \citep[blue circles;][]{Chen2025b}, and the SMC \citep[purple pentagons;][]{Dempsey}. CNM components in the LMC in the vicinity of the 30 Dor complex are additionally marked with yellow crosses. The dashed horizontal line indicates the $3\sigma$ sensitivity in optical depth (0.51) of the LGLBS IC~10 data. Our current sensitivity samples only the most opaque CNM structures, with most of the Magellanic Clouds' CNM components in the Magellanic Clouds falling below this threshold.

We use SMC data from \cite{Dempsey} to have a robust number of CNM components. To that end, we note that since no radiative transfer calculations were applied, we derive the FWHM $\Delta v_0$ using the provided equivalent width of the absorption spectra ($\int \tau dv$). We assume that each absorption spectrum has a Gaussian line profile and that it can be represented by a single component. While a single Gaussian function is a simplification, (since an absorption profile may consist of multiple overlapping CNM components), we note that \citet{Chen2025b} found that 80\% of the absorption spectra in the outskirts of the SMC were well described by a single Gaussian, making this a reasonable approximation. Therefore, the equivalent width can be treated as the area under the Gaussian curve, which can further be related to the FWHM $\Delta v_0$ and the peak optical depth $\tau_{\rm{peak}}$ as:

\begin{equation}\label{eq:fwhm}
    \text{FWHM }{\Delta v_0} = \frac{\int \tau(v) d v}{1.065 \times \tau_{\rm{peak}}}
\end{equation}

In Figure~\ref{fig:gp} we can see the CNM components in NGC~6822 all display narrow lines ($\Delta v_0 < 5$~km~s$^{-1}$), spanning a wide range of optical depths ($0.4 < \tau_{\rm peak} < 3.4$). In the SMC, most sources also exhibit narrow absorption features ($\Delta v_0 < 5$~km~s$^{-1}$), with only a few broader profiles ($\Delta v_0 \sim 7$–10~km~s$^{-1}$) associated with higher optical depths ($\tau_{\rm peak} > 2$).

By contrast, the IC~10 components show significantly larger linewidths ($\Delta v_0 \sim 13.6$~km~s$^{-1}$), exceeding what is expected from purely thermal broadening ($\sim 2$~km~s$^{-1}$) and surpassing those measured in the SMC \citep{Chen2025a}. In the LMC, most CNM components also have $\Delta v_0 < 6$~km~s$^{-1}$, but a substantial subset exhibits broader features with no clear correlation between linewidth and optical depth. Even within the 30~Doradus region, where CNM components trace intense star formation, linewidths range from $<5$~km~s$^{-1}$ to over 20~km~s$^{-1}$, showing no preferred regime.

Although the scatter in the FWHM $\Delta v_0$–$\tau_{\rm peak}$ plane appears largely random, \citet{Chen2025b} found that CNM components in the LMC tend to be broader along sightlines intersecting H~I shells, likely reflecting enhanced turbulence from stellar feedback and shell expansion. In our sample, only one source lies toward an H~I shell (\textsc{WBBH J002027+591728.7}), and its CNM component is indeed relatively broad, consistent with this trend. However, because two additional sightlines outside this shell region show one broad and one narrow component respectively, the interpretation remains uncertain. Moreover, since star-forming regions also coincide with high-density environments, enhanced cooling and fragmentation could produce a more clumpy CNM, further complicating the interpretation.
% reported systematically broader CNM components near H~I shells in the LMC. Such broadening may result from increased turbulence driven by stellar feedback and shell expansion. However, since star-forming regions also coincide with high-density environments, enhanced cooling and fragmentation could produce a more clumpy CNM, further complicating the interpretation.

\subsection{Fraction of Cold HI in IC10 and other dwarf galaxies}
Observationally, a global decreasing trend in $f_{\rm CNM}$ with metallicity is seen in absorption-line surveys of nearby low-metallicity dwarfs such as the SMC \citep{Dempsey} and LMC \citep{Chen2025b}, when considering mean $f_{\rm CNM}$ values averaged over hundreds of sightlines. 

We measure significantly higher CNM fractions in IC~10 per line of sight (20–33\%) compared to other low-metallicity galaxies. For instance, in the SMC’s inner body, \citet{Dempsey} estimated $f_{\text{CNM}} \approx 11\%$ using background sources and the relation $f_{\text{CNM}} \simeq T_c/\langle T_s\rangle$, assuming $T_c = 30$~K \citep[see][]{Dickey2000}. In the SMC outskirts, \citet{Chen2025a} found even lower values (1–11\%). In NGC~6822, \citet{nickngc} derived $f_{\text{CNM}} = 11.5$–$33\%$ based on absorption toward background point sources.  
These comparisons place IC~10 at the high end of cold gas fractions observed in low-metallicity dwarfs, with values closer to those measured for Galactic lines of sight \citep[median $\sim$30\%;][]{21spongea, naomi}.  

Theoretical predictions, such as those done with the TIGRESS-NCR numerical simulations \citep{kim2024ApJ...972...67K}, show that at low metallicity, the \textit{global} CNM fraction is expected to decrease, although the \textit{local} CNM mass fraction along individual lines of sight can remain similar. The balance between heating and cooling, which regulates thermal pressure equilibrium, is sensitive to the metallicity of the gas. When the dust-to-metal ratio is fixed (as is the case in the TIGRESS-NCR simulations except for the lowest metallicity models), the effects of the metallicity in metal cooling and grain photoelectric heating cancel out to the first order. In low metallicity environments, however, the reduced dust attenuation allows far-ultraviolet (FUV) radiation to penetrate more deeply, enhancing photoelectric heating for a given star formation rate. This results in more heating-dominated conditions at lower metallicities and, consequently, lower CNM area-filling factors.

% In low metallicity environments, however, the reduced dust attenuation facilitates penetration of FUV radiation, which in turn enhances photoelectric heating given star formation rate. This leads to more heating dominant conditions at lower metallicities and hence lower CNM area filling factors.
The elevated $f_{\text{CNM}}$ in IC~10 likely reflects an observational bias: our sightlines intersect dense environments near GMCs and H~II regions, where both the H~I column density and CNM fraction are naturally enhanced. This interpretation is consistent with Galactic studies showing higher $f_{\text{CNM}}$ near GMCs compared to diffuse sightlines (33\% vs.\ 22\%; \citealt{Stanimi2014, Nguyen2019}).  
In the LMC, \citet{Chen2025b} found that $f_{\rm CNM}$ spans 10–100\% (median 21\%, mean 27\%), also highlighting substantial spatial variation within a single galaxy. Our small sample of \textit{local} line-of-sight measurements is not enough to draw conclusions about the global $f_{\rm CNM}$ trends in IC~10. Future observations with larger samples will be essential to distinguish local from global dependencies of the CNM fraction on metallicity and environment.

\subsection{Line Blending}\label{sec:lineblend}
In Section~\ref{sec:line}, we examined whether the spatial coincidence of HI absorption with regions of intense star formation could explain broad FWHM ($\Delta v_0$) measured for \textsc{WBBH J002027+591706.1} and \textsc{WBBH J002017+591839.7}, drawing a comparison with CNM components in 30 Dor of the LMC. However, those LMC components show no systematic increase in linewidth despite residing in an extreme star-forming environment. In what follows, we explore alternative explanations for the observed broadening, including the influence of resolution effects and the relationship between H I absorption and molecular tracers.

Other than a turbulent, feedback-driven environment such as that of IC~10's star forming regions, resolution constraints likely contribute to the observed linewidths. 
Cold HI is expected to be highly clumpy on scales far smaller than our beam. Simulations suggest CNM forms at $\sim$0.1 pc scales, set by the WNM cooling length \citep{Hennebelle2007}. Observations in the Milky Way support this picture: CNM filaments have widths below 0.3 pc, in some cases as small as 0.09 pc \citep{Kalberla2016}\footnote{This estimate is based on HI emission data only, which has its own caveats.}.

This clumpiness parallels that observed in molecular gas. The CNM envelopes molecular clouds \citep{Stanimi2014}, which themselves are highly fragmented. High-resolution studies of low-metallicity galaxies such as the SMC show hierarchical molecular cloud fragmentation from ~9pc down to sub-parsec scales \citep[e.g.,][]{Jameson2018_CII, Tokuda2021, Saldano2023, Saldano2024}, with comparable 2pc-scale structure reported in NGC6822 and WLM \citep{Schruba2017, Rubio2015}, highlighting that molecular structures on scales from $<2$ pc down to sub-parsec are ubiquitous and likely mirrored in the cold HI.

Absorption studies probe CNM at higher effective resolution, but even on $\sim$9 pc scales blending and beam dillution limit what can be resolved \citep{Chen2025a}. At our $\sim$15 pc resolution, HI measurements inevitably average over many clumps, producing blended profiles.

\section{Conclusions}
We have presented new HI emission and absorption observations of the low-metallicity Local Group starburst dwarf galaxy IC~10, obtained by the LGLBS collaboration using the VLA. 
We have detected HI absorption in the direction of three radio continuum sources, these are among the first such detections in a low-metallicity star-forming environment beyond the Magellanic Clouds.\textit{The three detections occur towards internal HII regions of IC 10.} This effectively provides a direct probe of the densest cold gas embedded within IC~10’s star-forming regions rather than diffuse sightlines through the galaxy. 

We summarize our analysis and results below:
\begin{itemize}[nosep]
\item We identified CNM components along three separate lines of sight in IC~10 using Gaussian phase decomposition and radiative transfer. The spin temperature $T_s$ of these components range from $\sim 30 - 54$ K with a mean value of $44$K. We find column densities for the CNM ranging from $7.8 \times 10^{20}$ to $3 \times 10^{21}$ cm$^{-2}$ and $\Delta v_{\text{FWHM}}$ ranging from 5.7 km s$^{-1}$ to 13.6 km s$^{-1}$, corresponding to maximum $T_k$ ranging from 681 to 4032 K; and cold HI gas fraction $f_{\text{CNM}}$ from 0.20 to 0.38.

\item We measure cold gas fractions in IC~10 that are higher than those reported for the SMC and NGC 6822. This is likely influenced by our source selection, which is biased toward star-forming regions.
Future studies of different LGLBS targets will be critical for disentangling the relative roles of local star-forming environments and galaxy-wide metallicity in shaping the CNM fraction.

\item The HI absorption and molecular line emission (CO, HCO$^+$, HCN) in IC~10 are found at similar radial velocities, and their spectral features have comparable linewidths, indicating a well-mixed atomic and molecular component along these sightlines, and a denser CNM.

\item We find HI absorption in IC~10 with linewidths comparable to the CO, HCO$^+$, and HCN emission features, i.e. for \textsc{WBBH J002027+591706.1} the HI absorption feature has a FWHM of $13.60 \pm 0.78$ km s$^{-1}$ while the HCO$^+$ emission line has a FWHM of $11.0 \pm 0.6$ km s$^{-1}$. This indicates a clear kinematic relation that extends into the high-density ISM, consistent with models predicting higher CNM densities in low-metallicity conditions, though we note that selection biases may contribute to this trend.

\item For IC~10, a spatial resolution of $\sim$15 pc is insufficient to resolve individual CNM clouds. In addition, the absorption spectra were obtained against slightly extended HII regions internal to the galaxy. As a result, much of the observed line broadening could be attributed to spatial and kinematic averaging across extended background sources.

\end{itemize}

\begin{acknowledgments}
We thank the referee for their careful reading of the manuscript and for their helpful comments that improved the clarity and quality of this work.
I.A.S. thanks Dr. Trey Wenger for stimulating discussions on Gaussian fitting and radiative transfer.
This research was supported by the National Science Foundation awards 2205631, 2205630, 2205629, 2205631 and 2205630. The National Radio Astronomy Observatory and Green Bank Observatory are facilities of the National Science Foundation (NSF) operated under cooperative agreement by Associated Universities, Inc. V.V. acknowledges support from the Comité ESO Mixto 2024 and from the ANID BASAL project FB210003. This research made use of \texttt{astropy}, a community-developed core Python package for Astronomy \citep{astropy-1,astropy-2}. The data were imaged using CHTC services—Center for High Throughput Computing (2006): doi:10.21231/GNT1-HW21. 

\end{acknowledgments}
\bibliographystyle{aasjournal}
\bibliography{sample631}

\label{lastpage}

\end{document}